\documentclass[
superscriptaddress,
groupedaddress,
reprint,
nofootinbib,
amsmath,amssymb,
aps,
prd,
floatfix,
]{revtex4-2}
\usepackage[utf8]{inputenc}
\usepackage{graphicx}
\usepackage{dcolumn}
\usepackage{bm}
\usepackage{hyperref}
\usepackage{tensor}

\hypersetup{
    colorlinks=true,
    allcolors=blue,
}

\newcommand{\eq}[1]{Eq.~\eqref{#1}}

\newcommand{\ep}{\mathbf{e}_+}
\newcommand{\ec}{\mathbf{e}_\times}
\newcommand{\ex}{\mathbf{e}_x}
\newcommand{\ey}{\mathbf{e}_y}
\newcommand{\eb}{\mathbf{e}_b}
\newcommand{\el}{\mathbf{e}_l}

\usepackage{xcolor}

\begin{document}

\title{Reference Frames and Gravitational-Wave Polarizations: Symmetry Classification and Preferred-Frame Phenomenology}
\author{Jie Zhu}
 \email{jiezhu@cqu.edu.cn}
 \affiliation{Department of Physics and Chongqing Key Laboratory for Strongly Coupled Physics, Chongqing University, Chongqing 401331, P.R. China}

\author{Hao Li}
  \email{Corresponding author: haolee@cqu.edu.cn}
   \affiliation{Department of Physics and Chongqing Key Laboratory for Strongly Coupled Physics, Chongqing University, Chongqing 401331, P.R. China}

\date{\today}

\begin{abstract}
Gravitational wave (GW) polarizations are traditionally classified in a fixed frame ($E(2)$ classification), which does not account for how polarization patterns change under Lorentz boosts.
In this work, we derive the explicit transformation laws for all six GW polarizations under longitudinal and transverse boosts.
For gravity theories devoid of preferred frames, we propose a symmetry-based classification of the GW polarizations they admit. Among our key findings, we demonstrate that a propagating mode with five degrees of freedom strictly locks its longitudinal and breathing scalar amplitudes via the universal relation $A_l/A_b = -2(1-k^2/\omega^2)$.
For theories with a preferred frame, we analyze Bumblebee gravity and reveal that preferred-frame effects induce significant GW birefringence and observer-dependent polarization mixing.
Crucially, we identify a novel vector-to-tensor polarization conversion mechanism, where vector modes in the preferred frame inevitably generate observable tensor polarizations for moving detectors, offering a new pathway to test Lorentz-violating gravity.
Our framework provides a novel, observer‑independent classification of GW polarizations and reveals previously unnoticed polarization mixing effects.
\end{abstract}

\maketitle

\section{Introduction}

The direct detection of gravitational waves (GWs)~\cite{LIGOScientific:2016aoc, LIGOScientific:2018mvr, LIGOScientific:2020ibl, LIGOScientific:2021usb, KAGRA:2021vkt, LIGOScientific:2025hdt, LIGOScientific:2026sit} has opened a new window into the strong‑field dynamics of gravity and provides an unprecedented laboratory for testing the fundamental nature of spacetime.
Within the framework of General Relativity (GR), a massless spin-2 graviton manifests physically through merely two tensor polarization modes—the plus ($+$) and cross ($\times$) modes—both propagating strictly at the speed of light. However, generic modifications to Einstein's gravity, often motivated by high-energy physics or cosmological puzzles, naturally extend the radiative spectrum of gravity. 
Different theories of gravity predict different sets of GW polarizations—beyond the two tensor modes of GR, vector and scalar polarizations can appear in modified gravity frameworks~\cite{Hou:2017bqj, Jacobson:2004ts, Sagi:2010ei, Gong:2018vbo, Wagle:2019mdq, Bombacigno:2019did, Lu:2020eux, Dong:2021jtd, Farrugia:2018gyz, Soudi:2018dhv, Capozziello:2019msc, Capozziello:2020vil, Bahamonde:2021dqn, Capozziello:2021bki, Tachinami:2021jnf, Liang:2022hxd}. 
A correct identification of the polarization content is therefore essential for discriminating between GR and alternative theories.

The traditional classification of GW polarizations is formulated within the framework of the little group $E(2)$ of null rays~\cite{Eardley:1973br, Eardley:1973zuo}.
In this approach, one studies the Newman‑Penrose scalars associated with a plane GW propagating along a fixed spatial direction, and classifies the possible polarization states according to the non‑vanishing curvature components. A metric theory can, in principle, admit up to six independent polarizations: two tensor, two vector, and two scalar (breathing and longitudinal) polarizations.
This classification is complete for a given observer, but it is intrinsically tied to a fixed reference frame. It contains no information about how the observed polarization pattern changes when the observer moves relativistically with respect to the source or to the GW propagation direction. As GW detectors become more sensitive and future space‑based observatories (LISA, Taiji, TianQin) come online, the possibility of observing the same GW event from different inertial frames—or from moving detectors—motivates a systematic study of polarization transformations under frame changes.

In this work, we study the transformation behavior of GW polarizations under frame transformations and obtain the explicit formulae. Based on these results, we propose a polarization classification method utilizing kinematic symmetries for theories without a preferred frame. 
We find that a pure tensor mode must propagate at the speed of light; a pure vector mode is forbidden; scalar modes satisfy an invariant relation $A_l / A_b = 1-k^2/\omega^2$.
We present the first discovery that for a propagating mode with five degrees of freedom (e.g., a massive spin-2 mode), the breathing and longitudinal amplitudes are locked by a fixed relation $A_l / A_b = -2(1-k^2/\omega^2)$. For preferred-frame gravities, we take the Bumblebee vector-tensor theory as an example to illustrate the polarization mixing caused by motion relative to the preferred frame, and compare it with the Einstein-Aether theory. In particular, we show that the vector-to-tensor polarization conversion is prominent, identifying a new mechanism to detect vector polarization modes.

The remainder of this paper is organized as follows. In Sec.~\ref{sec:GWIntro}, we introduce the synchronous gauge framework for the six GW polarizations. 
In Sec.~\ref{sec:GWLT}, we explicitly derive the Lorentz transformation and polarization mixing laws under longitudinal and transverse boosts. 
Section~\ref{sec:noPF} delivers the complete symmetry-based classification for theories lacking preferred-frame effects. 
Section~\ref{sec:GWBumblebee} explores the preferred-frame phenomenology, polarization conversion, and birefringence in Bumblebee gravity, along with a detailed comparison to Einstein-Aether theory. 
A summary and final discussion are presented in Sec.~\ref{sec:summary}.

\section{Basics of GW Polarizations}\label{sec:GWIntro}


The physical manifestation of a GW polarization is the tidal deformation induced on a congruence of freely falling test particles. Therefore, GW polarizations are fundamentally observable quantities encoded in the linearized Riemann tensor rather than in the metric perturbation itself.
Under the assumption that the matter is not coupled with other fields, the observable effects of GWs are manifested in the geodesic deviation equation
\begin{equation}
\frac{d^2L^i}{dt^2}=-R^{(1)}_{0i0j}L^j. \label{eq:GDE}
\end{equation}
The polarization modes of gravitational waves are based on the relative motion of particles; therefore, all polarization information is contained within $R^{(1)}_{0i0j}$.
Without loss of generality, we assume the gravitational waves propagate along the $+z$ direction.
We choose to write the components of $R_{0i0j}$ in the following way to define the six polarization modes of GWs
\begin{equation}
R^{(1)}_{0i0j}
=\begin{pmatrix}
P_+ +P_b & P_\times & P_x\\
\\P_\times & -P_+ + P_b &P_y\\
\\P_x & P_y & P_l\end{pmatrix},
\end{equation}
where $P_+$ and $P_\times$ are tensor polarizations, $P_x$ and $P_y$ are vector polarizations, $P_l$ is the scalar longitudinal polarization, and $P_b$ is the scalar breathing polarization.

\begin{figure*}[htb]
	\centering
	\includegraphics[width=0.8\textwidth]{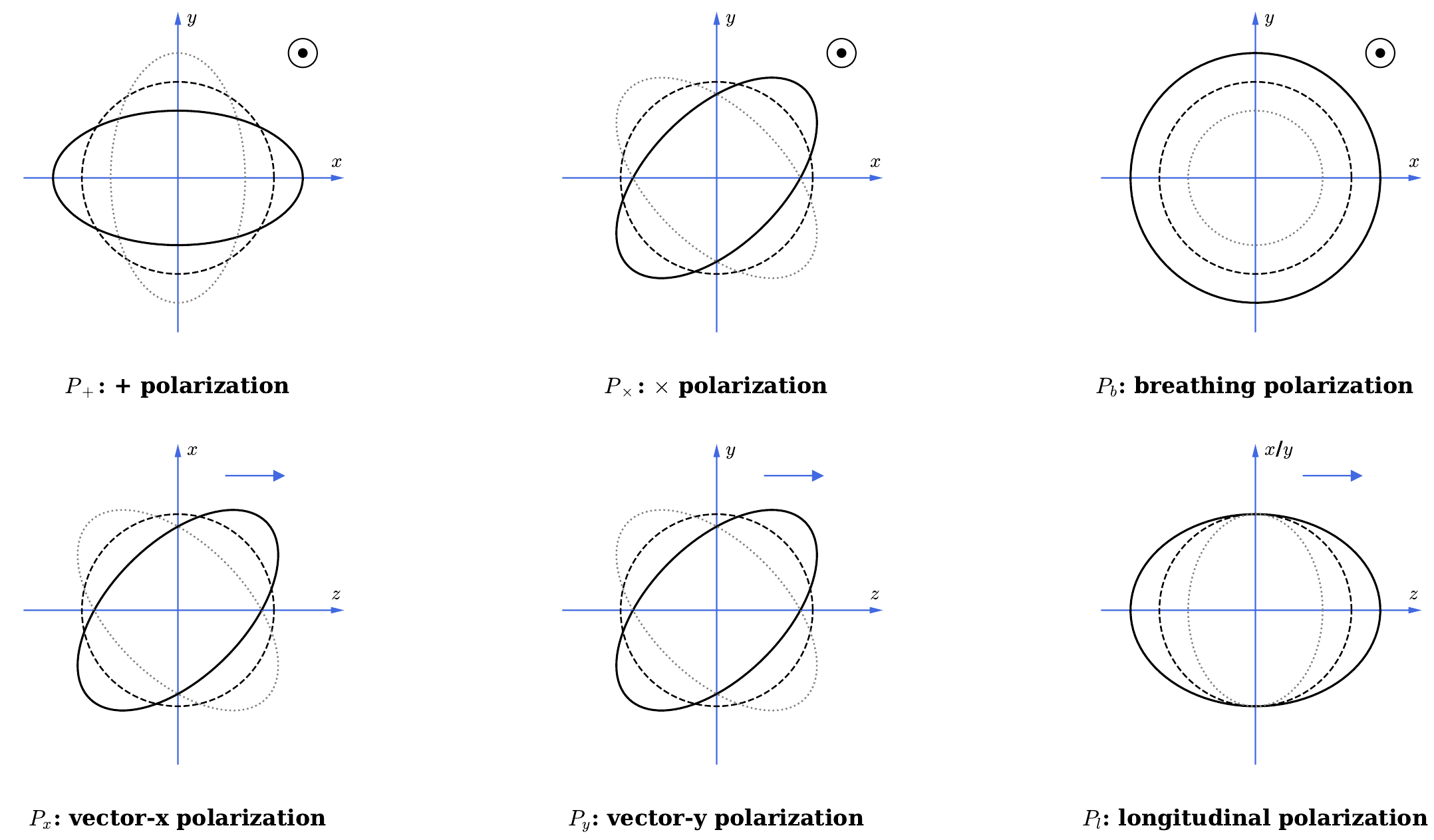}
	\caption{Six polarization modes of gravitational waves}
	\label{fig:polarizations}
\end{figure*}

The classification of GW polarizations is traditionally formulated within the framework of the little group $E(2)$ of null rays~\cite{Eardley:1973br, Eardley:1973zuo}. 
Although the $E(2)$ classification provides a complete taxonomy of GW polarizations in a given frame, it is intrinsically formulated with respect to a fixed observer.
Consequently, the standard $E(2)$ framework contains no information regarding the behavior of GW polarizations under changes of reference frame.
To address this issue, we investigate GW polarizations directly from the perspective of reference-frame transformations.

In GW analysis, gauge-invariant variables are useful for identifying the physical propagating degrees of freedom~\cite{Flanagan:2005yc}. 
However, their transformation properties under Lorentz boosts are generally obscure, making them unsuitable for a direct analysis of polarization mixing between different reference frames.
Therefore, we adopt the metric perturbation $h_{\mu\nu}$ of the gravitational field to investigate this problem directly.
At the linear level, diffeomorphism invariance manifests as a gauge transformation of the perturbation $h_{\mu\nu}$ generated by $\xi^\mu$ as
\begin{equation}
    h_{\mu\nu} \to h_{\mu\nu}-\partial_\mu \xi_\nu-\partial_\nu \xi_\mu.\label{eq:gauge}
\end{equation}
The metric perturbation $h_{\mu\nu}$ possesses 10 components, 4 of which are eliminated by the gauge redundancy, leaving 6 independent degrees of freedom that correspond to the 6 polarizations of the GW.
Therefore, in order to meaningfully compare gravitational waves between different reference frames, it is mandatory to fix the gauge. Here, we adopt the synchronous gauge condition, $h_{0\mu} = 0$, in which we have
\begin{equation}
R^{(1)}_{0i0j} = -\frac{1}{2}\ddot{h}_{ij}.
\end{equation}
We introduce the basis denoted by
\begin{equation}
\begin{aligned}
&\mathbf{e}_+ = 
\begin{pmatrix}1 & 0 & 0\\0 & -1 &0\\0 & 0 & 0\end{pmatrix},
\quad
&\mathbf{e}_\times = 
\begin{pmatrix}0 & 1 & 0\\1 & 0 &0\\0 & 0 & 0\end{pmatrix},
\\
&\mathbf{e}_x = 
\begin{pmatrix}0 & 0 & 1\\0 & 0 &0\\1 & 0 & 0\end{pmatrix},
\quad
&\mathbf{e}_y = 
\begin{pmatrix}0 & 0 & 0\\0 & 0 & 1\\0 & 1 & 0\end{pmatrix},
\\
&\mathbf{e}_b = 
\begin{pmatrix}1 & 0 & 0\\0 & 1 &0\\0 & 0 & 0\end{pmatrix},
\quad
&\mathbf{e}_l = 
\begin{pmatrix}0 & 0 & 0\\0 & 0 &0\\0 & 0 & 1\end{pmatrix},
\end{aligned}
\end{equation}
under which $h_{ij}$ can be decomposed as
\begin{equation}
h_{ij}=h_+ \mathbf{e}_+ +h_\times \mathbf{e}_\times
+h_x \mathbf{e}_x + h_y \mathbf{e}_y
+h_b \mathbf{e}_b + h_l \mathbf{e}_l,\label{eq:hij}
\end{equation}
where $h_+$, $h_\times$, $h_x$, $h_y$, $h_b$, and $h_l$ are functions of $t$ and $z$.
Under this decomposition, the relation between $R_{0i0j}$ and $h_{ij}$ can be expressed as
\begin{equation}
P_a = -\frac{1}{2}\ddot{h}_a(t,z),
\end{equation}
where the index $a$ denotes any of the polarization modes.
For a plane-wave ansatz of the form $h_a = A_a e^{i k_\mu x^\mu}$, we have 
\begin{equation}
    P_a = \frac{1}{2}\omega^2A_a e^{i k_\mu x^\mu}, \label{eq:Rh}
\end{equation}
where $\omega$ is the angular frequency of the GW.
This illustrates a prominent merit of the synchronous gauge framework: the geodesic deviation profile $P_a$ maps linearly onto the respective gravitational wave amplitude $h_a$. Consequently, the tensor components of $h_a$ serve as an excellent proxy to feature the actual physical observables.
For the gravitational waves propagate along the $+z$ direction, we assume that each polarization is
\begin{equation}
h_a = A_a \exp\left(i \Phi\right),\quad \Phi=\omega t- k z.\label{eq:waves}
\end{equation}

Throughout this work, we restrict ourselves to gravitational theories satisfying the following assumptions:
\begin{itemize}
    \item Gravity is described by the spacetime metric, which, in the vicinity of flat spacetime, admits the linear approximation $g_{\mu\nu} = \eta_{\mu\nu} + h_{\mu\nu}$;
    \item Test particles propagate along the geodesics of the spacetime metric, and consequently, their geodesic deviation is governed by Eq.~\eqref{eq:GDE};
    \item The theory satisfies diffeomorphism invariance. Alternatively, 
    at the linear level, the action possesses the gauge symmetry $h_{\mu\nu} \rightarrow h_{\mu\nu} -\partial_\mu\xi_\nu -\partial_\nu\xi_\mu$, so that the synchronous gauge can always be imposed.
\end{itemize}
Under these assumptions, the problem of GW polarizations can be reformulated as the problem of how the amplitudes of the six polarization components transform under changes of inertial reference frame. This will be the focus of the following section.

\section{Lorentz Transformation of GW} \label{sec:GWLT}

Since the action of the gravity considered here is invariant under diffeomorphisms, the linearized action is Lorentz invariant in the Minkowski background, and thus the metric perturbation $h_{\mu\nu}$ transforms as a rank-2 tensor.
This means that to perform a reference frame transformation, one needs to reconstruct the full metric perturbation $h_{\mu\nu}$ from all the modes, followed by applying a covariant Lorentz transformation to $h_{\mu\nu}$.
However, the gravitational wave in the new frame will generally violate the synchronous gauge condition. To ensure consistency, a supplementary gauge transformation must be performed after the frame transformation to restore the metric perturbation into the synchronous gauge form.
In addition, when the frame transformation is not aligned with the wave's propagation direction, the transformed gravitational wave will deviate from the $z$-axis. We must therefore perform a coordinate rotation to ensure that the wave propagates along the new $z$-axis.
Because any relative velocity can be decomposed into parts parallel and transverse to the wave propagation, without loss of generality, we can always choose coordinates where the transverse motion lies along the $x$-axis. Consequently, our analysis will focus on two representative cases: a Lorentz boost along the $z$-direction and a Lorentz boost along the $x$-direction.

\subsection{Transformation along the z Axis}\label{sec:trans-z}

Let the new coordinate system $S'$ be moving along the $z$-axis with a velocity $v_z$ relative to the original frame $S$.
In this frame transformation, the coordinate transforms as $x'^\mu = \Lambda^\mu{}_\nu x^\nu$.
Explicitly, it is
\begin{equation}
\begin{aligned}
    t' &= \gamma (t-v_z z),\\
    x' &=x,\\
    y' &=y,\\
    z' &= \gamma (-v_z t + z),
\end{aligned}
\end{equation}
where $\gamma = \frac{1}{\sqrt{1-v_z^2}}$.
The metric perturbation $h_{\mu\nu}$ transforms as
\begin{equation}
\begin{aligned}
h_{\mu\nu}^{\prime}(x^{\prime})
&=\frac{\partial x^\rho}{\partial x^{\prime\mu}}\frac{\partial x^\sigma}{\partial x^{\prime\nu}}h_{\rho\sigma}(x)\\
&=(\Lambda^{-1})^\rho{}_{\mu}\left(\Lambda^{-1}\right)^\sigma{}_{\nu}h_{\rho\sigma}{\left(\Lambda^{-1}x^{\prime}\right)}.
\end{aligned}
\end{equation}
As a consequence, the phase of a GW is invariant under the frame translations, and we have
\begin{equation}
\Phi '(x') = \Phi(x) = \Phi(\Lambda^{-1}x^{\prime}).
\end{equation}
Explicitly, it is
\begin{equation}
\begin{aligned}
\Phi'(x')&=\omega' t' -k' z' = \omega t - k z \\
&= \gamma (\omega-v_z k)t'-\gamma (k- v_z \omega)z'.
\end{aligned}
\end{equation}
As expected, the wave four-vector $(\omega, 0, 0, -k)$ transforms as a 1-form field.

For a GW present as equations~(\ref{eq:hij}) and (\ref{eq:waves}) in the frame $S$, the Lorentz-transformed fields in the frame $S'$ take the form $h'(x')_{\mu\nu} = C_{\mu\nu}\exp\left(i\Phi'(x')\right)$, where
\begin{equation}
C_{\mu\nu}
=\begin{pmatrix}
(\gamma^2-1) A_l & \gamma v_z A_x  & \gamma v_z A_y & \gamma^2 v_z A_l \\\\
\gamma v_z A_x & A_b +A_+  & A_\times & \gamma A_x\\\\
\gamma v_z A_y & A_\times  & A_b-A_+ & \gamma A_y\\\\
\gamma^2 v_z A_l & \gamma A_x & \gamma A_y  & \gamma^2 A_l
\end{pmatrix}.
\end{equation}
As we can see, the GW in the new frame $S'$ violates the synchronous gauge condition.
Now we perform the gauge transformation~(\ref{eq:gauge}) to restore $h'_{\mu\nu}$ to the synchronous gauge form.
The gauge parameters are chosen to be $\xi_\mu = C_\mu \exp(i\Phi'(x'))$ to eliminate the $0\mu$-components, where $C_\mu$ are constants.
The result is $h^{\rm syn \prime}_{i j} =A'_{ij} \exp\left(i\Phi'(x')\right)$, where
\begin{equation}
\begin{aligned}
A'_{ij} = 
&A_+ \ep' + A_\times \ec' +A_b \eb' \\
&+\frac{\omega}{\omega'}(A_x \ex' + A_y \ey')
+\frac{\omega^2}{\omega'^2}A_l \el',
\label{eq:newGWz}
\end{aligned}
\end{equation}
where $\omega'=\gamma(\omega-v_z k)$ is the angular frequency of the gravitational wave observed in the boosted frame $S'$, and the prime on the basis tensors represents the corresponding basis in $S'$.

As an alternative approach to restore the synchronous gauge, one can first calculate the Riemann tensor $R_{\mu\nu\rho\sigma}^{(1)}$ from the following relation
\begin{equation}
   R_{\mu\nu\rho\sigma}^{(1)} = \frac{1}{2} \left( \partial_\nu \partial_\rho h_{\mu\sigma} + \partial_\mu \partial_\sigma h_{\nu\rho} - \partial_\mu \partial_\rho h_{\nu\sigma} - \partial_\nu \partial_\sigma h_{\mu\rho} \right),
\end{equation}
and then obtain the components of $R_{0i0j}^{(1)}$, and finally apply the relation in Eq.~\eqref{eq:Rh} to derive $h^{syn \prime}_{ij}$. The final results confirm that both methods are entirely consistent.
Concurrently, if we first compute the Riemann tensor in the frame $S$, transform it to the frame $S'$ according to the standard 4-tensor transformation law, and subsequently utilize Eq.~\eqref{eq:Rh} to extract the synchronous gauge metric perturbations $h^{syn \prime}_{ij}$ in $S'$, the calculation yields an identical outcome. These non-trivial consistencies firmly demonstrate the correctness and robustness of our results.

We can clearly see that under a longitudinal boost along the $z$-axis, the distinct polarizations remain decoupled. 
The boost simply introduces a Doppler shift (manifested within $\Phi'(x')$) and modifies the amplitudes of these polarizations.
Remarkably, the plus, cross, and breathing polarizations exhibit amplitude invariance under the longitudinal boost.

Additionally, it is worth noting that the transformed amplitudes exhibit a pole at $v_z = \omega/k$, where $\omega'=0$. Given that $\omega/k$ is the phase velocity of the gravitational wave, a subluminal phase velocity ($v_p < 1$) implies the existence of a specific boost frame where the vector and longitudinal scalar polarizations diverge instantly. 
This divergence, however, is a mere mathematical artifact arising from the gauge-fixing nature of the synchronous gauge under a co-moving boost. The genuine, gauge-invariant physical observables are uniquely captured by the Riemann curvature tensor components $R_{0i0j}$, which remain perfectly regular.
Nevertheless, in the co-moving frame $S'$ where $\omega'=0$, the only non-vanishing component of the Riemann curvature tensor is $R_{0303}$, which reduces to
\begin{equation}
    R^{(1)}_{0303} = \frac{1}{2}\omega^2 A_l \exp\left(-i\sqrt{k^2-\omega^2}z\right),
\end{equation}
and becomes strictly static in time.
Although the Riemann tensor remains non-vanishing, the perturbation ceases to represent a propagating wave in the frame $S'$ since all temporal oscillations disappear. The configuration therefore represents a static tidal field rather than a propagating GW.
To avoid this situation, we need to require $\omega \ge k$.

\subsection{Transformation along the x Axis}

Now we consider the new coordinate system $S'$ moving along the $x$-axis with a velocity $v_x$ relative to the original frame $S$.
In this case, the phase of the GW transforms as
\begin{equation}
\Phi'(x') = \omega' t' - \vec{k}'\cdot \vec{x}' = \gamma \omega t' +\gamma v_x \omega x' - k z',
\end{equation}
where $\frac{1}{\sqrt{1-v_x^2}}$.
Under this boost, the propagation direction of the gravitational wave deviates from the new $z$-axis. Consequently, we need to define a new frame $S''$ that undergoes a coordinate rotation relative to $S'$.
We choose the rotation such that
\begin{equation}
\begin{aligned}
x'&= \frac{1}{\sqrt{k^2+\omega^2(\gamma^2-1) }}(k x''- \gamma v_x \omega z''),\\
y'&=y'',\\
z'&=\frac{1}{\sqrt{k^2+\omega^2(\gamma^2-1) }}(\gamma v_x \omega x''+ k z''),
\end{aligned}
\end{equation}
under which the phase becomes
\begin{equation}
\Phi''(x'') =
\omega'' t'' - k'' z''
=\gamma \omega t'' - \sqrt{k^2+\omega^2(\gamma^2-1) } z''.
\end{equation}
In the process of transforming $h_{\mu\nu}$ to the frame $S''$, we first perform a Lorentz boost followed by a spatial rotation, amounting to two successive Lorentz transformations. The resulting transformed field is given by $h''(x'')_{\mu\nu}=A''_{\mu\nu} \exp\left(i\Phi''(x'')\right)$, where
\begin{widetext}
\begin{equation}
A''_{\mu\nu}
=\begin{pmatrix}
\left(\gamma ^2-1\right) \left(A_b+A_+\right) 
&
\frac{\gamma ^2 v_x \left(k \left(A_b+A_+\right)+v_x\omega  A_x \right)}{\sqrt{\left(\gamma ^2-1\right) \omega ^2+k^2}} 
& 
\gamma  v_x A_\times
& 
\frac{\gamma  v_x \left(k A_x-\gamma ^2 v_x \omega  \left(A_b+A_+\right) \right)}{\sqrt{\left(\gamma ^2-1\right) \omega ^2+k^2}} 
\\\\
\frac{\gamma ^2 v_x \left(k \left(A_b+A_+\right)+v_x\omega  A_x \right)}{\sqrt{\left(\gamma ^2-1\right) \omega ^2+k^2}} 
& 
A_{11}
& 
\frac{\gamma  \left(k A_\times +v_x\omega  A_y \right)}{\sqrt{\left(\gamma^2-1\right) \omega ^2+k^2}} 
& 
A_{13}
\\\\

\gamma  v_x A_\times
& 
\frac{\gamma  \left(A_\times k+v_x\omega  A_y \right)}{\sqrt{\left(\gamma ^2-1\right) \omega ^2+k^2}} 
& 
A_b-A_+ 
& 
\frac{k A_y- \gamma ^2 v_x \omega  A_\times}{\sqrt{\left(\gamma ^2-1\right) \omega ^2+k^2}} 
\\\\
\frac{\gamma  v_x \left(k A_x-\gamma ^2 v_x\omega  \left(A_b+A_+\right) \right)}{\sqrt{\left(\gamma ^2-1\right) \omega ^2+k^2}} 
& 
A_{31}
& 
\frac{k A_y- \gamma ^2 v_x \omega  A_\times}{\sqrt{\left(\gamma ^2-1\right) \omega ^2+k^2}} 
& 
A_{33} 
\end{pmatrix},
\end{equation}
and
\begin{equation}
\begin{aligned}
A_{11} &= \frac{\gamma ^2 k^2( A_b + A_+) +\omega  \left(2 \gamma ^2 v_x k A_x +\left(\gamma ^2-1\right) \omega  A_l\right)}{\left(\gamma ^2-1\right) \omega ^2+k^2},
\\
A_{13} &=A_{31}= \frac{\gamma v_x  k \omega  \left(A_l-\gamma ^2 \left(A_b+A_+\right)\right)+\gamma\left(k^2-\left(\gamma ^2-1\right) \omega^2\right)  A_x }{\left(\gamma ^2-1\right) \omega ^2+k^2},
\\
A_{33} &= \frac{\gamma ^2 \omega  \left(\left(\gamma ^2-1\right) \omega  \left(A_b+A_+\right)-2v_x k A_x \right)+k^2 A_l}{\left(\gamma ^2-1\right) \omega ^2+k^2}.
\end{aligned}
\end{equation}
Next, we perform the gauge transformation~(\ref{eq:gauge}) to restore $h''_{\mu\nu}$ to the synchronous gauge form.
The result is shown to be $h^{\rm syn \prime\prime}_{i j} =A''_{ij} \exp\left(i\Phi''(x'')\right)$, where $A''_{ij} = A''_+ \ep''+A''_\times \ec''+A''_x \ex''+A''_y \ey'' +A''_b \eb''+A''_l \el''$, and
\begin{equation}
\begin{aligned}
A''_+ &=\frac{\left(\gamma ^2-1\right)\left(k^2-\omega ^2\right) A_b 
+\left(\left(\gamma ^2-1\right) \omega ^2+\left(\gamma ^2+1\right) k^2\right)A_+ 
+\omega  \left(2 \gamma ^2 v_x k A_x+\left(\gamma ^2-1\right) \omega  A_l\right)}{2 \left(\left(\gamma ^2-1\right) \omega ^2+k^2\right)},
\\
A''_\times &= \frac{\gamma  \left(A_\times k+\omega  A_y v_x\right)}{\sqrt{\left(\gamma ^2-1\right) \omega ^2+k^2}},
\\
A''_x &= \frac{\gamma ^2 k v_x \left(\left(k^2-\omega ^2\right)(A_+ + A_b)
+\omega ^2 A_l\right)+\omega   \left(\left(2 \gamma ^2-1\right) k^2-\left(\gamma^2-1\right) \omega ^2\right)A_x}{\gamma  \omega  \left(\left(\gamma ^2-1\right) \omega ^2+k^2\right)},
\\
A''_y &= \frac{ v_x(k^2-\omega^2)A_\times +k \omega  A_y}{\omega  \sqrt{\left(\gamma ^2-1\right) \omega ^2+k^2}},
\\
A''_b &= \frac{\left(\left(\gamma ^2-1\right) \omega ^2+\left(\gamma ^2+1\right) k^2\right)A_b 
+ \left(\gamma ^2-1\right) (k^2-\omega^2 ) A_+
+\omega  \left(2 \gamma ^2  v_x k A_x +\left(\gamma ^2-1\right) \omega  A_l\right)}{2 \left(\left(\gamma ^2-1\right) \omega ^2+k^2\right)},
\\
A''_l &= \frac{\left(\gamma ^2-1\right)  \left(k^2-\omega ^2\right)^2A_b
+ \left(\gamma ^2-1\right) \left(k^2-\omega ^2\right)^2A_+
+\gamma ^2 k \omega  \left(2 v_x \left(k^2-\omega^2\right)A_x+k \omega  A_l\right)}{\gamma ^2 \omega ^2 \left(\left(\gamma ^2-1\right) \omega ^2+k^2\right)}.\label{eq:newGW}
\end{aligned}
\end{equation}
\end{widetext}
Evidently, under a transverse boost along the $x$-direction, the distinct polarization modes in the $S''$ system are expressed as linear combinations of the modes from the $S$ system. Consequently, it is clear that a gravitational wave's polarization mode is not a frame-independent quantity.

Furthermore, we verify our results using another method: we compute the Riemann tensor $R^{(1)}_{\mu\nu\rho\sigma}$ in the $S$ frame, transform it to the $S''$ frame via two Lorentz transformations according to the four-tensor transformation laws, and then obtain the synchronous metric perturbation using Eq.~\eqref{eq:Rh}. The results show that these two methods yield identical outcomes, thereby verifying the reliability of our results.

When the gravitational wave propagates at the speed of light, we have $\omega = k$. Under this condition, the transformed polarization modes simplify to
\begin{equation}
\begin{aligned}
A''_{ij} = &
\left(A_{+} +v_x A_x +\frac{1}{2}v_x^2 A_l\right)\ep \\
&+(A_\times + v_x A_y)\ec \\
&+\frac{1}{\gamma}(A_x + v_x A_l)\ex
+\frac{A_y}{\gamma}\ey\\
&
+\left(A_{b} +v_x A_x +\frac{1}{2}v_x^2 A_l\right)\eb 
+\frac{A_l}{\gamma^2}\el.
\end{aligned}
\end{equation}
Under this light-speed condition, it is apparent that while the vector and longitudinal scalar modes mix into alternative polarization channels in the transformed frame, the plus and cross tensor modes, as well as the breathing scalar mode, do not generate cross-contributions. They exhibit only their own unique components, with their amplitudes being strictly preserved during the transformation.
Combining this with the previous results for the longitudinal transformation along the $z$-axis, we can conclude that for a gravitational wave propagating at the speed of light, its two tensor modes ($+$, $\times$) and the breathing scalar mode remain strictly invariant under arbitrary reference frame transformations. The sole effect of the Lorentz transformation is the Doppler frequency shift encoded within the phase.

\section{GW Polarizations in Gravity Theories Lacking Preferred-Frame Effects}\label{sec:noPF}

Here, we consider general theories of gravity from the perspective of reference frame transformations of gravitational wave modes. Before proceeding to the detailed analysis, we first clarify that although gravitational waves carry information regarding the underlying degrees of freedom (DoFs) of a given gravity theory, the dynamical degrees of freedom of gravity are conceptually and physically distinct from the gravitational wave polarization modes.
A classic example is a massive scalar-tensor theory~\cite{Hou:2017bqj}. While such theories possess only three propagating DoFs, they present four distinct GW polarization modes: two luminal tensor modes and two massive scalar modes. Crucially, the true massive scalar DoF is not identical to either scalar mode individually; rather, it is manifested through a specific mixture of both.

So it is essential to establish a precise nomenclature for the theoretical entities used in this work. We distinguish between a propagating mode, a propagating DoF, and a GW polarization based on a clear hierarchical structure:
\begin{itemize}
    \item A \textbf{Propagating Mode} is defined as an irreducible sector of the gravitational field characterized by its spin and mass within the framework of Lorentz invariant field theories (e.g., the massless spin-2 mode, the massive spin-2 mode, or the scalar mode). The individual components of a given propagating mode are governed by a single, common dispersion relation.
    \item A \textbf{Propagating degrees of freedom} refers to an individual dynamical component nested within a propagating mode; for example, the massive spin-2 mode contains exactly five independent propagating DoFs.
    \item A \textbf{Gravitational Wave Polarization} is strictly defined as the actual physical observable, manifesting as the specific pattern of tidal forces acting on test masses across reference frames.
\end{itemize}

We adopt the core principles of special relativity, which dictate that physical laws remain invariant across arbitrary reference frames. 
First, a propagating mode corresponds to an irreducible representation of the Lorentz group, and thus maintains its identity under any reference frame transformation. 
Because physical laws are identical in all frames, the underlying rules governing the GW polarizations contained within a specific propagating mode must likewise remain invariant. 
Consequently, if we assume that the polarizations of a GW constituting a given propagating mode satisfy a particular relation in a certain reference frame, this specific relation must hold true in any other frame as well. 
This relationship can manifest either qualitatively as the types and number of polarizations or quantitatively as the explicit relations satisfied by the amplitudes and wavevectors.

Following this guideline, we proceed to categorize the permissible configurations of GW polarizations within a specific propagating mode for Lorentz-invariant theories. Based on these underlying modes, we then systematically construct a full taxonomy of all potential polarization states.


\subsection{Tensor Polarizations Only}\label{sec:TO}

We first consider the scenario where the propagating mode exclusively comprises the tensor polarizations of the gravitational wave.
Owing to the fact that the two tensor polarizations are irreducibly mixed under a spatial rotation, this configuration implies that the corresponding propagating mode possesses a dynamical lower bound of at least two propagating DoFs.
Assuming that in the frame $S$, the GW in this propagating mode is 
\begin{equation}
h_{ij} = (A_{+}\ep + A_\times \ec) \exp(i \Phi(x)),  
\end{equation}
where $\Phi(x) = \omega t - k z$.
From \eq{eq:newGW}, we see that if $\omega \neq k$, then after a boost along the $x$-axis, vector and scalar polarizations emerge in the new reference frame. However, because this propagating mode is inherently frame-independent, it must exclusively contain tensor polarizations within the $S''$ frame as well.
Furthermore, as previously mentioned, when $\omega = k$, the tensor modes remain strictly invariant under reference frame transformations.
Consequently, we arrive at the following conclusion: 
\begin{itemize}
    \item In gravity theories free of preferred-frame effects, if a propagating mode exclusively comprises tensor polarizations, the gravitational waves associated with this propagating mode must propagate strictly at the speed of light. The amplitude of the tensor polarization mode remains invariant under reference frame transformations.
\end{itemize}
In this scenario, the two tensor polarizations represent a massless spin-2 mode with a helicity of $\pm 2$.
Indeed, GR serves as the primary example here, featuring only a massless spin-2 propagating mode that manifests physically through just two tensor polarizations.

\subsection{Vector Polarizations Only}\label{sec:VO}

Here, we consider the scenario where the propagating mode only contains vector polarizations.
In other words, in the frame $S$, the GW in this propagating mode is
\begin{equation}
h_{ij} = (A_x\ex + A_y \ey) \exp(i \Phi(x)),
\end{equation}
where $\Phi(x) = \omega t - k z$.
From \eq{eq:newGW}, we see that in the reference frame $S''$, the gravitational wave generically contains both tensor and scalar polarizations; demanding the vanishing of these tensor and scalar polarizations yields the trivial constraint $A_x = A_y = 0$.
Consequently, we arrive at the following conclusion: 
\begin{itemize}
    \item In gravity theories free of preferred-frame effects, there exists no propagating mode that exclusively comprises vector polarizations.
    If a propagating mode contains vector polarizations, the gravitational waves associated with this propagating mode must simultaneously carry both tensor and scalar polarizations.
\end{itemize}


\subsection{Scalar Polarizations Only}\label{sec:SO}

We then consider the case in which the propagating mode only contains scalar polarizations.
In the frame $S$, the GW in this propagating mode is
\begin{equation}
h_{ij} = (A_b\eb + A_l \el) \exp(i \Phi(x)),
\end{equation}
where $\Phi(x) = \omega t - k z$.
From \eq{eq:newGW}, we see that in the frame $S''$, the GW transforms to $h^{\rm syn \prime\prime}_{i j} =A''_{ij} \exp\left(i\Phi''(x'')\right)$, where
\begin{equation}
\begin{aligned}
A''_{ij} = &
\frac{\left(\gamma ^2-1\right) \left(A_b \left(k^2-\omega ^2\right)+\omega ^2 A_l\right)}{2 \left(\left(\gamma ^2-1\right) \omega ^2+k^2\right)}
\ep''
\\&+
\frac{\gamma  k v_x \left(A_b \left(k^2-\omega ^2\right)+\omega ^2 A_l\right)}{\omega  \left(\left(\gamma ^2-1\right) \omega ^2+k^2\right)}
\ex''
\\&+
\frac{ \left(\left(\gamma ^2-1\right) \omega ^2+\left(\gamma ^2+1\right) k^2\right)A_b+\left(\gamma ^2-1\right) \omega ^2 A_l}{2 \left(\left(\gamma ^2-1\right) \omega^2+k^2\right)}
\eb''
\\&+
\frac{\left(\gamma ^2-1\right)  \left(k^2-\omega ^2\right)^2A_b +\gamma ^2 k^2 \omega ^2 A_l}{\gamma ^2 \omega ^2 \left(\left(\gamma ^2-1\right) \omega ^2+k^2\right)}
\el''.
\end{aligned}
\end{equation}
It is evident that if
\begin{equation}
A_l = \left(1-\frac{k^2}{\omega^2}\right)A_b,\label{eq:lb}
\end{equation}
then only the two scalar polarizations persist in the $S''$ frame, with all tensor and vector polarizations identically vanishing.
In this condition, we have
\begin{equation}
A_{ij} = A_b \eb + \left(1-\frac{k^2}{\omega^2}\right)A_b\el, \label{eq:scalarForm}
\end{equation} and
\begin{equation}
\begin{aligned}
A''_{ij} = & A_b \eb + \frac{\omega^2-k^2}{\gamma^2\omega^2}A_b \el\\
=& A_b \eb + \left(1-\frac{k''^2}{\omega''^2}\right)A_b\el,
\end{aligned}
\end{equation}
where
\begin{equation}
\omega '' = \gamma \omega, \quad
k'' = \sqrt{k^2+\omega^2(\gamma^2-1) }
\end{equation}
are precisely the transformed components of the four-wavevector $(\omega, 0, 0, -k)$ within the $S''$ frame.
As is demonstrated, upon undergoing a coordinate transformation along the $x$-direction, the GW of this particular combination preserves its purely scalar profile without changing its structural form.
Additionally, from \eq{eq:newGWz}, we can verify that if \eq{eq:lb} is satisfied, the transformed GW remains strictly form-invariant under \eq{eq:scalarForm} even after a boost along the $z$-axis.
If this GW travels at the speed of light, then we have $\omega=k$, and consequently $A_l=0$ and only the breathing polarization survives.
Therefore, for a propagating mode restricted solely to scalar GW polarizations, we arrive at the following conclusions:
\begin{itemize}
    \item If the wave propagates strictly at the speed of light, this propagating mode of the GW only contains the breathing polarization, and the amplitude of the breathing mode remains invariant under reference frame transformations.
    \item If the propagation speed is non-luminal, there exists only a single propagating DoF. Its associated GW simultaneously exhibits both the breathing and longitudinal polarizations, where the ratio of their amplitudes in any arbitrary reference frame is uniquely determined by the wavevector as \eq{eq:lb}; furthermore, the amplitude of the breathing polarization remains invariant under reference frame transformations.
\end{itemize}

Indeed, this state of affairs is already present in certain modified gravity models, most notably Horndeski theory~\cite{Hou:2017bqj}. Here, the theory contains three propagating DoFs, combining a massless tensor mode with a scalar mode. When the scalar mode is massless, it gives rise solely to the breathing polarization; whereas a massive scalar mode inevitably generates a mixture of breathing and longitudinal polarizations.

Another case in point is $f(R)$ gravity, which features four polarization modes: two standard tensor polarizations moving at the speed of light, alongside two scalar polarizations~\cite{RizwanaKausar:2016zgi}. 
Nevertheless, advanced analysis confirms that the theory contains just three propagating DoFs, as the two scalar polarizations intertwine to form a single scalar propagating mode possessing only one individual DoF~\cite{Liang:2017ahj}.

A meticulous derivation demonstrates that the amplitude ratio of the breathing to longitudinal polarizations under the synchronous gauge matches \eq{eq:lb} identically for both theoretical frameworks. 
Moreover, within the massive scalar sector, the energy-momentum relation enforces $\omega^2 = k^2 + m^2$ (with $m$ being the mass of the scalar mode), under which condition \eq{eq:lb} reduces identically to $A_l/A_b=m^2/\omega^2$.
For the sake of brevity, we refrain from displaying the extensive algebraic details of this calculation here.

\subsection{Mixing of Two Polarizations}\label{sec:2P}

Here, we consider the scenario where a propagating mode simultaneously comprises two distinct gravitational wave polarizations. This implies that the propagating mode must maintain this specific polarization configuration invariant under any reference frame transformation.

For a gravitational wave containing exclusively tensor and vector polarizations in the $S$ frame, \eq{eq:newGW} dictates that the breathing mode in the $S''$ frame vanishes if and only if $A_x = 0$. However, since we can always select the boost direction such that the component of the vector mode along this specific boost direction is non-zero, it follows that a self-consistent propagating DoF cannot consist solely of these two polarizations.

The remaining two cases are entirely analogous: a mixture of vector and scalar polarizations will inevitably generate tensor polarizations in the boosted frame, whereas a combination of scalar and tensor polarizations will give rise to vector polarizations in the new reference frame.
Therefore, we arrive at the following conclusion:
\begin{itemize}
    \item The GW associated with a propagating mode cannot consist exclusively of exactly two categories of polarizations (i.e., tensor-vector, tensor-scalar, or vector-scalar hybrids).
\end{itemize}

\subsection{Mixing of Three Polarizations}\label{sec:3P}

Finally, we turn our attention to the scenario where a propagating mode simultaneously comprises a mixture of all three categories of polarizations.

Firstly, the case where there are six propagating DoFs and each exhibits all six polarizations is trivially satisfied. Because the entire six-dimensional polarization space is already spanned, any reference frame transformation merely reshuffles these six modes without generating any external components.
Secondly, the number of polarization components contained in such a propagating mode must strictly exceed four. This is because, upon a coordinate rotation by $\theta$ around the $z$-axis, the vector polarizations rotate by $\theta$ while the tensor polarizations rotate by $2\theta$. This discrepancy implies that if such a state contains only a single vector component and a single tensor component in one frame, one can always find a rotated frame where both vector polarizations and both tensor polarizations are simultaneously excited.
Therefore, we are motivated to investigate whether a non-trivial scenario exists where a propagating mode accommodates five DoFs that span six gravitational wave polarizations. This particular configuration requires the two scalar polarizations to bind into a specific combination whose structure remains strictly invariant under reference frame transformations.

Analogous to the case of purely scalar modes, we consider the situation where the amplitudes of the two scalar modes satisfy the relation
\begin{equation}
A_l = G (\omega, k )A_b,
\end{equation}
where $G(\omega, k )$ is a function to be determined.
We require that the functional form of this relation remains strictly form-invariant under any frame transformation. Specifically, upon a Lorentz boost along the $x$-axis, we expect that in the $S''$ frame, we have
\begin{equation}
    A_l'' = G(\omega'', k'') A_b'',
\end{equation}
where
\begin{equation}
\omega '' = \gamma \omega, \quad
k'' = \sqrt{k^2+\omega^2(\gamma^2-1) }.
\end{equation}
We begin by considering an infinitesimal boost along the x-axis, namely, $v_x \ll 1$, which implies that the Lorentz factor expands as $\gamma = 1 + \mathcal{O}(v_x^2)$.
Under this infinitesimal transformation, we have
\begin{equation}
G(\omega'', k'') = G(\omega, k) + \mathcal{O}(v_x^2).
\end{equation}
On the other hand, from \eq{eq:newGW}, we have
\begin{equation}
\frac{A_l''}{A_b''}=G(\omega, k ) +\frac{(2k^2-2\omega^2-\omega^2 G(\omega, k))v_x A_x}{\omega k A_b} + \mathcal{O}(v_x^2).
\end{equation}
Therefore, demanding that \eq{eq:lbG} remains form-invariant under this infinitesimal transformation uniquely fixes the explicit expression $G(\omega,k)= -2 (1-k^2/\omega^2)$, and we arrive at the relation
\begin{equation}
A_l = -2\left(1-\frac{k^2}{\omega^2}\right) A_b. \label{eq:lbG}
\end{equation}
Although \eq{eq:lbG} remains form-invariant under infinitesimal transformations, we must verify whether this invariance persists under general, finite reference frame transformations.
From \eq{eq:newGW} and \eq{eq:newGWz}, one can easily verify that this relation remains invariant under reference frame transformations along both the $x$- and $z$-axes. Consequently, it follows that this relation is preserved under any arbitrary reference frame transformation.
In this scenario, a reference frame transformation along the $x$-axis causes these five DoFs to mix non-trivially with one another, such that none of the individual amplitudes remains invariant under the frame transformation.
Thus, we arrive at the following conclusion: 
\begin{itemize}
    \item If a gravitational theory devoid of preferred-frame effects admits a propagating mode comprising five degrees of freedom, then the amplitudes of the two scalar polarizations within this mode must satisfy Eq.~\eqref{eq:lbG}.
\end{itemize}
Simultaneously, we have the following conclusion:
\begin{itemize}
    \item If a specific propagating mode of a gravitational theory exhibits no scalar polarizations in a given reference frame but manifests non-vanishing scalar polarizations in another, then the amplitudes of the two scalar polarizations within the newly transformed frame satisfy Eq.~\eqref{eq:lbG}.
\end{itemize}

For this propagating mode, comprised of five propagating DoFs, a massive tensor mode serves as a canonical example. 
The most classic realization is massive gravity~\cite{Hinterbichler:2011tt, deRham:2014zqa}.
Although the standard Fierz-Pauli action~\cite{Fierz:1939ix} for massive gravity lacks gauge invariance and thus fails within our prescribed framework, the restoration of gauge symmetry via the Stueckelberg technique allows us to perform a rigorous analysis within the synchronous gauge. 
A meticulous evaluation demonstrates that the amplitudes of the two scalar gravitational wave modes precisely satisfy \eq{eq:lbG} (see Appendix~\ref{sec:GWPMG}).
In the context of the massive tensor mode, upon substituting the on-shell dispersion relation $\omega^2 = k^2 + m^2$, \eq{eq:lbG} further simplifies to $A_l/ A_b = -2 m^2/\omega^2$.

Such a massive tensor configuration is also naturally realized within higher-derivative gravity frameworks. A classic paradigm is quadratic gravity~\cite{Stelle:1977ry, Stelle:1976gc}, in which the foundational Hilbert-Einstein action is supplemented by the quadratic curvature invariants $R^2$ and $C_{\mu\nu\rho\sigma}C^{\mu\nu\rho\sigma}$, where $C_{\mu\nu\rho\sigma}$ is the Weyl tensor.
In quadratic gravity, in addition to the massless tensor mode and the massive scalar mode, there also exists a massive, ghost-like tensor mode~\cite{Stelle:1977ry, Stelle:1976gc, Zhu:2026jkf}.
The GW polarizations of the massless tensor and massive scalar propagating modes are in perfect alignment with the discussions in Sec.~\ref{sec:TO} and Sec.~\ref{sec:SO}, respectively.
Regarding the tensor mode, a comprehensive discussion was also delivered by Alves et al.~\cite{Alves:2022yea}, wherein the explicit polarization tensors of the gravitational waves were computed. By applying an appropriate gauge transformation to map their results into the synchronous gauge, we find that our foundational relation \eq{eq:lbG} is, remarkably, satisfied once again.

\subsection{Summary of Possible Polarizations}\label{sec:PolarizationsSummary}

For gravitational theories devoid of preferred-frame effects, by synthesizing the polarization analyses of the distinct propagating modes investigated above, we can comprehensively determine the permitted gravitational wave polarizations within such a theoretical framework. The definitive classification is summarized as follows:
\begin{enumerate}
    \item Two tensor polarizations ($+$ and $\times$) propagating at the speed of light. These polarizations correspond to the massless spin-2 graviton with helicity $\pm 2$, which are strictly transverse and traceless, satisfying the standard propagation properties of GR.
    
    \item A scalar breathing polarization propagating at the speed of light. This polarization stems from a massless spin-0 scalar field with helicity 0, inducing an isotropic, transverse deformation on the test particle array while leaving the longitudinal displacement entirely unaffected.
    
    \item A coupled combination of the breathing and longitudinal polarizations, with their amplitude ratio tightly constrained by $A_l / A_b = 1 - k^2/\omega^2$. A quintessential representative of this class is the massive scalar propagating mode, where the physical mass breaks the pure transversality and introduces a distinctive longitudinal polarization.
    
    \item A propagating mode possessing five physical degrees of freedom, yet manifesting as a comprehensive mixture of tensor, vector, and scalar polarizations, wherein the amplitudes of its scalar sectors strictly satisfy the interlinked relation $A_l / A_b = -2(1 - k^2/\omega^2)$. The quintessential paradigm for this scenario is the massive tensor propagating mode (as realized in Fierz-Pauli or quadratic gravity).
    
    \item The seemingly straightforward configuration featuring six distinct propagating DoFs paired with six independent polarizations. Intriguingly, no viable modified gravity framework discovered to date populates this sector, as the structural symmetry of the spacetime metric consistently intertwines the scalar components into the fine-tuned invariant relations derived above.
\end{enumerate}

It is worth discussing the limiting behavior of Case 4. When $k = \omega$, this mode propagates at the speed of light, under which the general scalar constraint reduces to $A_l / A_b = 0$. Consequently, the longitudinal scalar polarization completely vanishes, leaving a configuration composed of two tensor modes, two vector modes, and a single pure breathing polarization.
Nevertheless, despite its theoretical admissibility in our classification scheme, this specific light-speed polarization mixture finds no physical realization within established modified gravity theories to date.
This discrepancy highlights that while spacetime symmetries define the outer boundaries of what is theoretically possible, the detailed gauge dynamics of specific field theories consistently selectively populate these geometric channels.

Another crucial point worthy of discussion is that, across the entire taxonomy, vector polarizations manifest exclusively in Case 4 and the trivial Case 5. In both regimes, tensor and scalar polarizations invariably emerge in tandem. This implies that for any modified gravity theory devoid of preferred-frame effects, the presence of vector polarizations rigidly dictates the co-existence of both tensor and scalar sectors, all of which are governed by a unified dispersion relation identical to that of the vector mode.

Furthermore, the foundational premise of the discussion herein resides in the structural invariance of the underlying theory. Specifically, the analysis is formulated by demanding that the gravitational wave polarizations inherent to each dynamic propagating mode exhibit strict form-invariance across all relativistic reference frames. At its core, this approach constitutes a pristine symmetry-based analysis that bypasses model-dependent ambiguities.
In comparison, the effective field theory (EFT) method offers a complementary symmetry-based paradigm. Literature utilizing this technique establishes the scalar-tensor actions through derivative expansions and maps out the corresponding polarization taxonomy, providing a dynamical benchmark that beautifully cross-validates our geometric classification~\cite{Dong:2023bgt}.
In Ref.~\cite{Dong:2023bgt}, Dong et al. formulated four distinct Propositions regarding the gravitational wave polarizations within generic scalar-tensor theories featuring higher-derivative extensions. Crucially, their Propositions 1 and 4 directly correspond to specific projections of our tensor propagating modes, while the non-trivial amplitude ratio between the two scalar polarizations dictated by their Proposition 3 identically reproduces our Case 3. Granted, their bottom-up EFT approach successfully rules out the light-speed propagating vector modes—a dynamical elimination that remains outside the scope of our purely kinematic framework. Nevertheless, the operator-dependent nature of their methodology fails to fully map out the broader radiative landscape uncovered herein. Specifically, their Propositions 1 and 4 yield only fragmented insights into the massive tensor propagation, and more noticeably, their framework completely glosses over the invariant ratio of $A_l / A_b = -2m^2/\omega^2$ that uniquely characterizes the scalar sectors within our massive tensor configurations (Case 4).

\section{Preferred-Frame Effects and GW Polarizations in Bumblebee Gravity}\label{sec:GWBumblebee}

Building upon the invariant polarization taxonomy analyzed in the previous section, we now turn our attention to theories characterized by preferred-frame dependencies. Crucially, we preserve the Lorentz covariance of the gravitational action at the linear level, meaning that any preferred-frame artifact arises purely from the non-trivial background configurations of non-scalar sectors. A prime illustration is Bumblebee gravity~\cite{Kostelecky:2003fs, Bluhm:2004ep}. In this framework, the spontaneous symmetry breaking of a vector field yields a non-zero vacuum expectation value (VEV), which non-minimally couples to the spacetime curvature, offering a rigorous theoretical sandbox for investigating preferred-frame gravitational radiations.

Focusing on Bumblebee gravity, this section maps out the explicit gravitational wave polarizations altered by the presence of a preferred cosmic frame. We contrast these results directly with the preferred-frame-free classification established in the preceding sections, highlighting the severe distortion of the standard polarization relations. Crucially, we delineate the unique observational footprints that these modifications imprint on multi-messenger data streams.

\subsection{Basics of Bumblebee Gravity}

The action of the general bumblebee gravity is
\begin{equation}
\begin{aligned}
    S= \int d^4x\sqrt{-g}&\left(\frac{1}{2\kappa}\left(R+\lambda B_\mu B^\mu R + \xi B^\mu B^\nu R_{\mu\nu}\right)\right.\\
    &\left.-\frac{1}{4}B_{\mu\nu}B^{\mu\nu}-V\right)+S_\mathrm{m},  \label{eq:action}
\end{aligned}
\end{equation}
where $g$ is the determinant of the metric $g_{\mu\nu}$, the constant $\kappa\equiv8\pi G$ with $G$ being the gravitational constant, $S_{\mathrm{m}}$ represents the action for matter fields of no interest in this work, $B_\mu$ is the bumblebee field, and the field strength tensor is $B_{\mu\nu}=\partial_{\mu}B_{\nu}-\partial_{\nu}B_{\mu}$.
In bumblebee theories, the potential $V$ is selected to provide a non-vanishing VEV for $B_\mu$, and could have the following general functional form
\begin{equation*}
    V=V(X),\quad X\equiv B^\mu B_\mu + s b^2,\label{potential}
\end{equation*}
where $b$ is a positive real constant, and $s=\pm 1$ or 0 to determine whether the expection value of $B_\mu$ is timelike, spacelike or lightlike. 
In the literature, it is usually assumed that \(V\) has (at least one of) its minimum/maximum at \(0\), thus
\begin{equation*}
    V(0)=0,\ \text{and}\ V'(0)=0.\label{vacuumcondition}
\end{equation*}
The VEV of the bumblebee field is determined when $B^\mu B_\mu + s b^2=0$, and this equation provides a non-null vacuum expectation value
\begin{equation*}
    \langle B^\mu\rangle=b^\mu,
\end{equation*}
where $b_\mu b^\mu + s b^2=0$.

However, recent investigations~\cite{Zhu:2026hxm} indicate that from the perspective of Hamiltonian analysis, the spontaneous breaking of the vector field imposes significantly more stringent constraints on the functional form of the potential $V(X)$. 
Specifically, the Hamiltonian consistency conditions dictate that the background VEV must be strictly timelike or lightlike, and the potential must be at least cubic in $X$~\cite{Zhu:2026hxm}.
For the polarization analysis, we exploit two crucial theoretical leverage points. First, ensuring spatial isotropy constrains the background configuration to a purely timelike VEV, allowing us to choose a specific cosmic frame where $b^\mu \propto \delta^\mu_0$. Second, since the potential is structurally fixed to be of cubic or higher order, it drops out of the linearized field equations, exerting zero influence on the propagation modes and polarization profiles of the gravitational radiations.

It is noteworthy that the gravitational wave polarizations in bumblebee gravity with $\lambda=0$ and a quadratic potential have been previously investigated in Ref.~\cite{Liang:2022hxd}. 
Therein, the authors presented a highly convoluted, case-by-case taxonomy of polarization states; specifically, under the ansatz that the wave propagates along the $z$-axis, their findings indicated that the resulting polarization profiles heavily hinge upon whether the spatial $x$- and $y$-components of the background VEV vanish. However, this apparent structural complexity is kinematically transparent: a specific polarization mode evaluated in the pristine rest frame (where the background field possesses only the $t$-component) will naturally project out and contribute to additional polarization channels upon transforming to a new reference frame where the $x$- and $y$-components of the background field become non-vanishing. To explicitly substantiate this point, we shall first execute the gravitational wave polarization analysis within the pure temporary background frame, and subsequently elucidate how these decoupled physical modes systematically reshuffle into mixtures of distinct polarization profiles under general frame transformations.

Importantly, recent literature has illuminated that from the perspective of post-Newtonian approximations, the bumblebee gravity action admits viable parametrized post-Newtonian (PPN) solutions precisely when the coupling parameters adhere to the critical constraint $\lambda = -\xi/2$~\cite{Zhu:2026ikz}. 
Under this fine-tuned parameter space, the non-minimal curvature coupling collapses into the elegant form of $B_\mu B_\nu G^{\mu\nu}$. 
An identical restriction has concurrently emerged within the framework of recent cosmological perturbation theories~\cite{vandeBruck:2025aaa}. 
Motivated by these theoretical cross-validations, we partition our subsequent polarization analysis into two distinct regimes: $\lambda + \xi/2 = 0$ and $\lambda + \xi/2 \neq 0$. Crucially, we shall explicitly demonstrate that the dynamic degrees of freedom as well as the active gravitational wave polarization profiles differ fundamentally between these two parameter configurations.

\subsection{GW Polarizations in the Preferred-Frame}

The equation of motions for $g_{\mu\nu}$ is 
\begin{equation}
G_{\mu\nu}=\kappa T_{\mu\nu},
\end{equation}
where $T_{\mu\nu}=T_{\mu\nu}^{\mathrm{M}}+T_{\mu\nu}^{B}$, 
$T_{\mu\nu}^{\mathrm{M}}$ is the energy-momentum tensor of matter, and 
\begin{equation}
\begin{aligned}
T_{\mu\nu}^{B}=&
-B_{\mu\alpha}B_{\nu}^{\alpha}
-\frac{1}{4}B_{\alpha\beta}B^{\alpha\beta}g_{\mu\nu}
-Vg_{\mu\nu}+2V^{\prime}B_{\mu}B_{\nu}
\\&
+\frac{\xi}{\kappa}\left(
\frac{1}{2}B^{\alpha}B^{\beta}R_{\alpha\beta}g_{\mu\nu}
-B_{\mu}B^{\alpha}R_{\alpha\nu}
-B_{\nu}B^{\alpha}R_{\alpha\mu}\right.
\\&
+\frac{1}{2}\nabla_{\alpha}\nabla_{\mu}\left(B^{\alpha}B_{\nu}\right)
+\frac{1}{2}\nabla_{\alpha}\nabla_{\nu}\left(B^{\alpha}B_{\mu}\right)
\\&
\left.
-\frac{1}{2}\nabla^{2}\left(B_{\mu}B_{\nu}\right)
-\frac{1}{2}g_{\mu\nu}\nabla_{\alpha}\nabla_{\beta}\left(B^{\alpha}B^{\beta}\right)\right)
\\&
+\frac{\lambda}{\kappa}\Big(
-B_\alpha B^\alpha G_{\mu\nu}
-B_\mu B_\nu R
\\&
+2\nabla_\nu(B_\alpha \nabla_\mu B^\alpha)
-2g_{\mu\nu} \nabla_\alpha(B_\beta \nabla^\alpha B^\beta)\Big).
\end{aligned}
\end{equation}
For the $B_\mu$ sector, the equation of motion is
\begin{equation}
\nabla^{\mu}B_{\mu\nu}
+\frac{1}{\kappa}(\xi B^\mu R_{\mu \nu}+\lambda B_\nu R)-2V' B_\nu =0.\label{eq:EOMB}
\end{equation}
We linearize the equations of motion within the preferred reference frame characterized by the background field $b_\mu = (b, 0, 0, 0)$. Under the metric and vector field perturbations specified by $g_{\mu\nu} = \eta_{\mu\nu} + h_{\mu\nu}$ and $B_\mu = b_\mu + \tilde{b}_\mu$, we obtain
\begin{equation}
R^{(1)} = \partial^\mu \partial^\nu h_{\mu\nu} - \Box h,
\end{equation}
\begin{equation}
R^{(1)}_{\mu\nu} = \frac{1}{2}\left[
\partial_\mu \partial^\alpha h_{\alpha\nu} +\partial_\nu \partial^\alpha h_{\alpha\mu} 
- \partial_\mu \partial_\nu h
-\Box h_{\mu\nu}
\right],
\end{equation}
\begin{equation}
B^{(1)}_{\mu\nu} = \partial_\mu \tilde{b}_\nu - \partial_\nu \tilde{b}_\mu,
\end{equation}
\begin{equation}
V(x)^{(1)} = V'(X)^{(1)}=0.
\end{equation}
Following the approach outlined in Ref.~\cite{Jacobson:2004ts}, we perform the polarization analysis for the gravitational waves.
By inserting the relations into the complete field equations, the linearized equations of motion can be explicitly derived.
We can adopt the synchronous gauge to perform the GW polarization analysis, which fixes the gauge freedom by setting $h_{0i} = 0$.
We assume a GW of the form
\begin{equation}
\begin{aligned}
h_{ij} = \epsilon_{ij} \exp (i k_\mu x^\mu),\\
B_\mu = \epsilon_{i} \exp (i k_\mu x^\mu),
\end{aligned}
\end{equation}
and choose coordinates such that the wavevector is $(\omega,0,0,-k)$.
By inserting the wave solutions into the linearized field equations, the corresponding gravitational wave polarizations can be explicitly extracted. Herein, we define a \emph{propagating mode} as a set of solutions governed by an identical dispersion relation.

For the case $\lambda + \xi/2 \neq 0$, the gravitational wave solutions manifest as three propagating modes within the unique frame where the bumblebee field is restricted to its temporal component.
The first propagating mode consists exclusively of solutions containing tensor polarizations, and its dispersion relation is given by
\begin{equation}
\omega = v_{T} k, \quad v_{T}^2 = 1+\frac{\xi b^2}{1-(\xi+\lambda)b^2}.\label{eq:BumblebeeT}
\end{equation}
Owing to the linear nature of the dispersion relation, the characteristic velocity $v_{T}$ represents both the phase velocity and the group velocity. 
In this propagating mode, the polarization tensors $\epsilon_{ij}$ and $\epsilon_{i}$ satisfy the following constraint conditions:
\begin{equation}
\epsilon_i =0, \quad \epsilon_{i 3}=0, \quad \epsilon_{11} = -\epsilon_{22},
\end{equation}
with $\epsilon_{11}$ and $\epsilon_{12}$ serving as free parameters that represent the two physical DoFs.
Within this tensor mode, the perturbations of the background field $\tilde{b}_\mu$ do not propagate.

The second propagating mode consists exclusively of vector polarizations, and its dispersion relation is given by
\begin{equation}
\omega = v_{V} k, \quad v_{V}^2 = 1+\frac{\xi^2 b^2}{2\kappa (1-(\lambda+\xi)b^2)}.\label{eq:BumblebeeV}
\end{equation}
In this propagating mode, the polarization tensors $\epsilon_{ij}$ and $\epsilon_{i}$ satisfy the following constraint conditions:
\begin{equation}
\begin{gathered}
\epsilon_{11}=\epsilon_{12}=\epsilon_{21}=\epsilon_{22}=\epsilon_{33}=\epsilon_{0}=\epsilon_{3}=0,\\
\frac{\epsilon_1}{\epsilon_{13}}= \frac{1-(\lambda+\xi)b^2}{\xi b}v_{V},
\end{gathered}
\end{equation}
with $\epsilon_{13}$ and $\epsilon_{23}$ serving as free parameters that represent the two physical DoFs.
Within this propagating mode, the perturbations of the background field exhibit transverse propagation.

Furthermore, it is instructive to note that because $\xi b^2$ scales the magnitude of Lorentz symmetry breaking, under the parametric condition $\lambda \sim \xi$ alongside a weak Lorentz-violating background, the velocity approximates as $v_{V}^2 \simeq 1 + \frac{\xi^2 b^2}{2\kappa} > 1$. Consequently, this indicates that the vector modes exhibit superluminal propagation in this specific rest frame.
This outcome is not surprising; we have pointed out in Sec.~\ref{sec:trans-z} that the phase velocity for the vector polarizations is better to be superluminal.
If the dispersion relation of these vector modes were massive, this would be harmless because the group velocity remains subluminal. For bumblebee gravity, however, the relation is linear, making the phase and group velocities identical. Therefore, the consistency of the theory ensures that the propagation speed of the vector polarizations is superluminal.

The third propagating mode consists exclusively of scalar polarizations, and its dispersion relation is given by
\begin{equation}
\omega = v_S k, \quad v_S^2 = 1+\frac{\xi (2\xi-\kappa)b^2}{3\kappa (1-(\lambda+\xi)b^2)}.\label{eq:BumblebeeS}
\end{equation}
In this propagating mode, the polarization tensors $\epsilon_{ij}$ and $\epsilon_{i}$ satisfy the following constraint conditions:
\begin{equation}
\begin{gathered}
\epsilon_{12} = \epsilon_{13} = \epsilon_{23} = \epsilon_{2} = \epsilon_{3}=0, \\
\epsilon_{22}=\epsilon_{11},\\
\frac{\epsilon_{33}}{\epsilon_{11}} = -2(1-v_S^{-2}),\\
\frac{\epsilon_0}{\epsilon_{11}} = \frac{1-(\lambda+\xi)b^2}{(2\lambda+\xi)b},\\
\frac{\epsilon_3}{\epsilon_{11}} = -\left(\frac{\xi b}{\kappa}+\frac{1-(\lambda+\xi)b^2}{(2\lambda+\xi)b}\right)\frac{1}{v_S},
\end{gathered}
\end{equation}
with $\epsilon_{11}$ serving as one free parameter that represents the only physical DoF.
Within this propagating mode, the perturbations of the background field exhibit longitudinal propagation.
It is noteworthy that the amplitude ratio between the longitudinal and breathing polarizations here precisely satisfies Eq.~\eqref{eq:lbG}; that is to say, under any arbitrary reference frame transformation, the amplitude ratio within the transformed frame persistently obeys Eq.~\eqref{eq:lbG}.

We now turn to the critical case where $\lambda + \xi/2 = 0$. It is straightforward to see that while the generic parameter limits $\lambda \to -\xi/2$ for the tensor and vector polarizations are well-behaved, this limiting procedure induces a divergence in the scalar polarization sector. Consequently, one must impose $\lambda = -\xi/2$ directly within the full field equations prior to performing the gravitational wave polarization analysis. The final results reveal that, distinct from the standard $2\text{T}+2\text{V}+1\text{S}$ result, the case with $\lambda + \xi/2 = 0$ propagates only $2\text{T}+2\text{V}$ DoFs. The original scalar degree of freedom degenerates, yet the tensor and the vector polarizations remain fully consistent with those obtained via the smooth limit of the generic case.
Explicitly, in this specific parametric regime, the dispersion relations for the two tensor polarizations take the form
\begin{equation}
\omega = v_{T} k, \quad v_{T}^2 = 1+\frac{\xi b^2}{1-\xi b^2/2},
\end{equation}
and the polarization tensors $\epsilon_{ij}$ and $\epsilon_{i}$ satisfy the following constraint conditions:
\begin{equation}
\epsilon_i =0, \quad \epsilon_{i 3}=0, \quad \epsilon_{11} = -\epsilon_{22},
\end{equation}
with $\epsilon_{11}$ and $\epsilon_{12}$ serving as free parameters that represent the two physical DoFs.
The dispersion relations for the two vector polarizations take the form 
\begin{equation}
\omega = v_{V} k, \quad v_{V}^2 = 1+\frac{\xi^2 b^2}{2\kappa (1-\xi b^2/2)},
\end{equation}
and the polarization tensors $\epsilon_{ij}$ and $\epsilon_{i}$ satisfy the following constraint conditions:
\begin{equation}
\begin{gathered}
\epsilon_{11}=\epsilon_{12}=\epsilon_{21}=\epsilon_{22}=\epsilon_{33}=\epsilon_{0}=\epsilon_{3}=0,\\
\frac{\epsilon_1}{\epsilon_{13}}= \frac{1-\xi b^2/2}{\xi b}v_{V},
\end{gathered}
\end{equation}
with $\epsilon_{13}$ and $\epsilon_{23}$ serving as free parameters that represent the two physical DoFs.
This dispersion relation matches exactly with the result obtained within cosmological perturbation theory~\cite{vandeBruck:2025aaa}.


In the following, we shall proceed to consider the gravitational wave polarizations of Bumblebee gravity in a general reference frame. To this end, here we clarify why a theory seemingly characterized by Lorentz violation can nevertheless have its gravitational wave polarizations treated via standard Lorentz transformations.

This argument starts with the action Eq.~\eqref{eq:action}, which satisfies diffeomorphism invariance. When linearized around the background, the linear action is Lorentz invariant against a Minkowski background. Specifically, the linearized action remains invariant under the Lorentz transformations where $b_\mu$ and $\tilde{b}_\mu$ transform as vectors and $h_{\mu\nu}$ transforms as a rank-2 tensor.
Consequently, for the underlying equations of motion, if $\tilde{b}_\mu(x^\mu)$ and $h_{\mu\nu}(x^\mu)$ solve the system configured by the background $b_\mu$, then within the primed frame $S'$, the configuration given by $\tilde{b}'_\mu(x'^\mu)$ and $h'_{\mu\nu}(x'^\mu)$ is guaranteed to be a solution under the background $b'_\mu$. This holds true as long as the relevant fields adhere strictly to their corresponding Lorentz transformation laws.
Because gravitational waves are solutions to the linearized equations, their Lorentz-transformed forms remain solutions to the corresponding equations under the transformed background $b'_\mu$. Therefore, to obtain the gravitational wave polarizations under a generic background, we only need to find a specific Lorentz transformation that connects the preferred frame $S$ (where $b_\mu$ has only a time component) to the general frame, transform the wave solutions from the preferred frame, and analyze them there.

\subsection{GW Phase in General Frames}

We first consider the transformation of the GW phase. For a wave solution of the form $h_{\mu\nu} = A_{\mu\nu} \exp(i k_\mu x^\mu)$, its Lorentz transformation is given by
\begin{equation}
\begin{aligned}
h_{\mu\nu}^{\prime}(x^{\prime}) =& (\Lambda^{-1})^\rho{}_{\mu}\left(\Lambda^{-1}\right)^\sigma{}_{\nu}A_{\rho\sigma}\exp\left(i k_\alpha (\Lambda^{-1})^\alpha{}_\beta x^{\prime \beta}\right)\\
&\propto \exp\left(i k'_\beta x^{\prime \beta}\right),
\end{aligned}
\end{equation}
where $k'_\beta = (\Lambda^{-1})^\alpha{}_\beta k_\alpha$. 
Since the gauge transformations acting on GWs do not alter their phase, we arrive at the conclusion that the gravitational wavevector transforms strictly as a vector field under Lorentz transformations. 
Given that the gravitational wavevector obeys a linear dispersion relation in the frame $S$, it follows that the transformed wavevector in any arbitrary inertial reference frame will persistently satisfy a linear dispersion relation.


\begin{figure}[htb]
	\centering
	\includegraphics[width=0.8\linewidth]{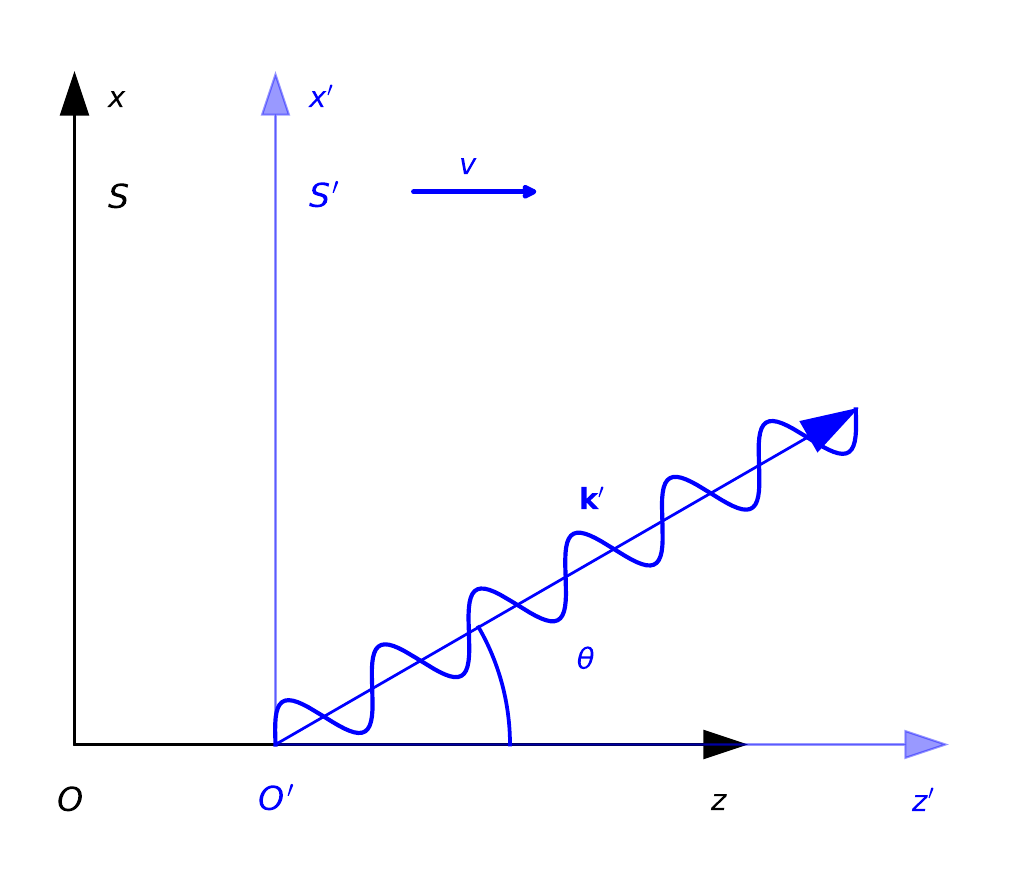}
	\caption{Schematic illustration of GW observations in the lab frame $S'$.}
	\label{fig:frames}
\end{figure}
We now consider our laboratory frame $S'$ moving at a velocity $v$ relative to the preferred frame $S$ (See Fig.~\ref{fig:frames}). We construct the coordinate system such that the relative velocity of $S'$ with respect to $S$ is given by $\vec{v} = (0, 0, v)$. Suppose a gravitational wave is observed in the $S'$ frame, propagating at an angle $\theta$ with respect to the $z'$-axis, so that its corresponding wavevector can be parametrized as $(\omega', k' \sin\theta', 0, k' \cos\theta')$, which satisfies the linear dispersion relation $\omega' = u' |\vec{k}'|$. Now, considering this gravitational wave within the preferred frame $S$, the corresponding wavevector is uniquely determined by a Lorentz boost along the $z$-axis. Within the $S$ frame, the dispersion relation remains strictly linear as $\omega = u |\vec{k}|$, where $u$ denotes the propagation velocity of the distinct mode derived in the preceding section. An explicit calculation demonstrates that
\begin{equation}
\begin{aligned}
    u' =& u \frac{1-v^2}{1-u^2 v^2}\sqrt{\cos^2\theta+\frac{1-u^2 v^2}{1-v^2}\sin^2\theta}\\
    &-v \frac{1-u^2}{1-u^2v^2}\cos\theta.\label{eq:uPrime}
\end{aligned}
\end{equation}
Evidently, when the propagation velocity $u$ within the preferred frame $S$ deviates from the speed of light, both the gravitational wave velocity and its corresponding dispersion relation in the laboratory frame $S'$ become direction-dependent. In particular, the wave velocity attains its minimum and maximum values at $\theta' = 0$ and $\theta' = \pi$, respectively, and the resulting wave velocity in the $S'$ frame falls precisely into the interval $[\frac{u-v}{1-uv},\frac{u+v}{1+uv}]$.

Now, we assume that the velocity $v$ of the laboratory frame $S'$ relative to the preferred frame $S$ is small, and that the deviation of the gravitational wave velocity from the speed of light in the $S$ frame, defined as $\delta u \equiv u-1$, is also a small quantity. Under these assumptions, the deviation of the wave velocity $u'$ from the speed of light within the $S'$ frame is given by
\begin{equation}
    \delta u' \simeq (1+2v \cos\theta) \delta u.
\end{equation}
The cosmological comoving frame, also known as the CMB rest frame, serves as a natural candidate for the preferred frame. In this configuration, the bumblebee vacuum expectation value is restricted to its temporal component. According to the Planck satellite data, the Earth moves at a speed of $v = 0.00123\;c$ relative to this CMB rest frame~\cite{Planck:2013kqc}.
Hence, it follows that $0.998 < \delta u' / \delta u < 1.002$, indicating that the intrinsic relative motion of the Earth has a minor kinematic impact on altering the gravitational wave speed.
Thus, while previous works constraining bumblebee gravity via gravitational wave speed ignored the effect of the Earth's relative motion~\cite{Lai:2025nyo}, this correction is negligible, and their conclusions remain reliable.
Based on the joint observations of GW170817 and its electromagnetic counterparts~\cite{LIGOScientific:2017zic}, we have $-3\times10^{-15}\leq \delta u'\leq7\times10^{-16}$.
Since the velocity deviation for the tensor modes satisfies $\delta u'\simeq\delta u \simeq \xi b^2/2$, the resulting constraint on bumblebee gravity remains $-6\times10^{-15}< \xi b^2<1.4\times10^{-15}$.

\subsection{Polarization Mixing in General Frames}

We next turn our attention to the polarization mixing of gravitational waves within the transformed reference frame $S'$.
Without loss of generality, here we focus on the parameter branch $\lambda \neq -\xi/2$. The alternative scenario, $\lambda = -\xi/2$, can be derived based on this by removing the scalar propagation DoF in the frame $S$ and evaluating the appropriate limit for the tensor and vector DoFs.

For generic parameters, the dispersion relations for the three propagation modes in Eqs.~\eqref{eq:BumblebeeT}, \eqref{eq:BumblebeeV}, and \eqref{eq:BumblebeeS} are distinct. Because their phases also differ after transforming to frame $S'$, we must process each mode individually.
As reference frame transformations along the propagation direction preserve the GW polarizations, we focus on the relative motion of frame $S'$ perpendicular to this direction, with the polarization transformation described by Eq.~\eqref{eq:newGW}.

Since the amplitude ratio of the two scalar polarizations within the $S$ frame satisfies the relation given by Eq.~\eqref{eq:lbG}, according to the discussion in Sec.~\ref{sec:3P}, we establish that this specific ratio remains universally locked by Eq.~\eqref{eq:lbG} across all three propagation modes in any arbitrary reference frame, owing to the geometric nature of the reference frame transformation. Specifically, if we denote $A'^{a}_{b}$ as the amplitude representing the conversion of the $a$-mode from the $S$ frame into the $b$-polarization within the $S'$ frame, we obtain
\begin{equation}
\begin{gathered}
    \frac{A^{\prime T}_l}{A^{\prime T}_b} = -2 (1-{v'_T}^{-2}),\\
    \frac{A^{\prime V}_l}{A^{\prime V}_b} = -2 (1-{v'_V}^{-2}),\\
    \frac{A^{\prime S}_l}{A^{\prime S}_b} = -2 (1-{v'_S}^{-2}),
\end{gathered}
\end{equation}
where $v'_T$, $v'_V$, and $v'_S$ denote the respective propagation velocities of the corresponding gravitational wave modes within the $S'$ frame, as uniquely determined by their specified directions of propagation.

Next, let us evaluate the approximate forms of the polarization amplitudes in the $S'$ frame, assuming that the departure of the wave velocity from light speed in the $S$ frame is highly suppressed, and that the transverse boost velocity $v_x$ remains non-relativistic. 
For a gravitational wave possessing tensor polarizations with amplitudes $A_+$ and $A_\times$ in the $S$ frame, we denote  $\delta v_T = v_T - 1$ as the tensor velocity deviation. According to Eq.~\eqref{eq:newGW}, up to the leading order, the corresponding polarization amplitudes observed within the transformed frame $S'$ are given by
\begin{equation}
\begin{aligned}
    A'_+ &= (1-v_x^2 \delta v_T) A_+,\\
    A'_\times &= (1-v_x^2 \delta v_T) A_\times,\\
    A'_x &= -2 v_x \delta v_T A_+,\\
    A'_y &= -2 v_x \delta v_T A_\times,\\
    A'_b &= -v_x^2 \delta v_T A_+,\\
    A'_l &= 4v_x^2\delta v_T^2 A_+.
\end{aligned}
\end{equation}
Therefore, we can see that the tensor polarizations in frame $S$ are suppressed when transforming into other components in frame $S'$. Specifically, the components transforming into vector, breathing, and longitudinal polarizations are suppressed by $v_x \delta v_T$, $v_x^2 \delta v_T$, and $v_x^2 \delta v_T^2$, respectively.

Next, we consider the transformation of the vector polarizations in frame $S$. Using $A_x$ and $A_y$ for the amplitudes in $S$, and defining $\delta v_V = v_V - 1$, the leading-order results of the transformation are
\begin{equation}
\begin{aligned}
    A'_+ &= v_x A_x,\\
    A'_\times &= v_x A_y,\\
    A'_x &= (1-\frac{1}{2}v_x^2) A_x,\\
    A'_y &= (1-\frac{1}{2}v_x^2) A_y,\\
    A'_b &= v_x A_x,\\
    A'_l &=  -4 v_x \delta v_V A_x.
\end{aligned}
\end{equation}
We can see that, except for the longitudinal polarization, which is suppressed by $v_x \delta v_V$ due to Eq.~\eqref{eq:lbG}, the amplitudes of the other components are proportional to $v_x$. Specifically, the vector polarization in $S$ contributes to the tensor polarization in $S'$ proportional to $v_x$. If the CMB rest frame is the preferred frame, the Earth's motion causes the observed tensor polarization to include a $0.001$-level vector polarization, providing a concrete test for future gravitational wave detectors.

Finally, we consider the different polarization components of the scalar modes in frame $S'$. Letting $A_b$ be the breathing mode amplitude in frame $S$, while the longitudinal mode is given by Eq.~\eqref{eq:lbG}, and defining $\delta v_S = v_S - 1$, we have
\begin{equation}
\begin{aligned}
    A'_+ &= -3 v_x^2\delta v_S A_b,\\
    A'_\times &=0,\\
    A'_x &= -6 v_x \delta v_S A_b,\\
    A'_y &=0, \\
    A'_b &= (1-3 v_x^2 \delta v_S)A_b,\\
    A'_l &= -4\delta v_S A_b.
\end{aligned}
\end{equation}
Thus, for the breathing polarization in frame $S$, its transformations into other polarization components in frame $S'$ are all suppressed.

\subsection{Birefringence in General Frames}

Based on the formulation established above, we now investigate the phenomenology of the birefringence phenomenon within general reference frames, induced by the existence of such a preferred frame.

Let us assume a scenario where a gravitational wave source and a distant observer are at rest in the preferred frame $S$. The observer will detect three distinct signals (two groups of signals if $\lambda = -\xi/2$) arriving at different times from the same direction, with each signal exhibiting a pure, unmixed polarization state.

In frame $S'$ moving relative to $S$, the wavevectors of the three signal groups point in different directions due to their inconsistent dispersion relations. Meanwhile, as each group generates other polarization components in frame $S'$, the observation in $S'$ shows that at \emph{three different times}, three signal groups are observed from \emph{three different directions}, each containing \emph{all six polarizations}. Among them, one group has the strongest tensor amplitude, one has the strongest vector amplitude, and one has the strongest scalar amplitude.

If the propagation velocities of the three polarizations are identical, their wavevectors are also the same, and the birefringence phenomenon disappears. Obviously, these three velocities coincide when $\xi = 2\kappa$, and the velocity is
\begin{equation}
v_V^2=v_T^2=v_S^2 = 1+\frac{2\kappa b^2}{1-(2\kappa+\lambda)b^2}.
\end{equation}
For a small Lorentz violation, this velocity is superluminal.
Notably, despite the preferred-frame effects in this gravitational theory, the polarization classification in this case perfectly matches Case 4 in Sec.~\ref{sec:PolarizationsSummary}.
The preferred-frame effect materializes in the primed frame $S'$ via the modified dispersion relation $\omega' = u' k'$, wherein the propagation speed $u'$ acquires a non-trivial angular dependence uniquely determined by Eq.~\eqref{eq:uPrime}.

\subsection{Comparing with Einstein-Aether Theory}

The Einstein-Aether theory is likewise a modified gravity framework endowed with a preferred reference frame~\cite{Jacobson:2000xp, Eling:2003rd, Jacobson:2004ts, Eling:2004dk, Foster:2005dk}, whose action is formulated as
\begin{equation}
    S=\frac{1}{16\pi G}\int\sqrt{-g}(R+L_{\mathrm{aether}}+L_{\mathrm{matter}})d^4x.
\end{equation}
The second term, the aether Lagrangian, is given by
\begin{equation}
    L_{\mathrm{aether}}=-K^{ab}{}_{mn}\nabla_au^m\nabla_bu^n+\lambda(g_{ab}u^au^b+1),
\end{equation}
where $K^{ab}{}_{mn}$ is defined as
\begin{equation}
    K^{ab}{}_{mn}=c_1g^{ab}g_{mn}+c_2\delta_m^a\delta_n^b+c_3\delta_n^a\delta_m^b-c_4u^au^bg_{mn},
\end{equation}
being the $c_i$ dimensionless coupling constants, and $\lambda$ a Lagrange multiplier enforcing the unit timelike constraint on the aether $u^a$.

It is a subtle yet crucial point that, although the Bumblebee gravity can mimic a subset of the Einstein-Aether theory when its potential is rigidly constrained via a Lagrange multiplier, these two frameworks harbor a fundamental conceptual distinction. Despite both being categorized as vector-tensor theories, the fundamental dynamical variables are specified differently: the core field in Einstein-Aether theory is defined as the contravariant vector $u^\mu$, whereas in the Bumblebee model, it is the covariant tensor $B_\mu$. This structural disparity in the choice of the primary independent variable leads to distinctly different variational derivatives with respect to the metric tensor. Consequently, even for actions that appear formally identical, they yield entirely different energy-momentum tensors, radically altering the underlying gravitational dynamics of the two theories.

We here consider the GW polarizations of the Einstein-Aether theory within the aether rest frame where $u^a = (1,0,0,0)$. Denoting $c_{ij\dots} \equiv c_i + c_j + \dots$, the gravitational radiation for generic parameters exhibits the standard $2\text{T}+2\text{V}+1\text{S}$ polarization structure~\cite{Jacobson:2004ts}. In the synchronous gauge, the dispersion relation for the pure tensor polarizations is formulated as
\begin{equation}
    \omega = v_T k,\quad v_T^2 = \frac{1}{1-c_{13}}.
\end{equation}
For the pure vector polarizations, the dispersion relation is
\begin{equation}
    \omega = v_T k,\quad v_V^2 = \frac{c_1-\frac{1}{2}c_1^2+\frac{1}{2}c_3^2}{c_{14}(1-c_{13})}.
\end{equation}
For the scalar polarization, the dispersion relation is
\begin{equation}
    \omega = v_S k,\quad v_S^2 = \frac{c_{123}(2-c_{14})}{c_{14}(1-c_{13})(2+2c_2+c_{123})},
\end{equation}
and the amplitude ratio between the longitudinal polarization and the breathing polarization is
\begin{equation}
    \frac{A_l}{A_b} = \frac{2[c_{123}^2-(1+c_2)(c_{24}-c_3)]}{c_{123}(2-c_{14})}.
\end{equation}
It is particularly noteworthy that, unlike in the Bumblebee gravity, the amplitude ratio of the scalar modes in the Einstein-Aether theory does not satisfy Eq.~\eqref{eq:lbG}. Consequently, the constraint relation Eq.~\eqref{eq:lbG} is by no means a universal feature applicable to general modified gravity frameworks endowed with a preferred reference frame.

Two distinct parametric branches merit a detailed investigation, the first of which is the decoupling limit $c_{13} = 0$. Under this boundary, although the aether field itself undergoes transverse excitations, the gravitational vector sector suppresses entirely since its radiative amplitude scales linearly with $c_{13}$. Furthermore, because the tensor speed collapses back to unity ($v_T = 1$), the tensor amplitudes exhibit rigid invariance under arbitrary frame operations. However, because the scalar amplitude ratio violates both Eq.~\eqref{eq:lbG} and Eq.~\eqref{eq:lb} in this regime, a non-trivial kinematic consequence emerges: when observed from a generically transformed frame, the underlying scalar channel will project out supplementary, apparent vector and tensor modes as frame-induced artifacts.

The second special parametric regime of interest is defined by the condition $c_{14} = 0$. Under this configuration, both the vector and scalar polarization sectors vanish identically, leaving the two standard tensor degrees of freedom within the aether rest frame. However, as long as the propagation velocity of this remaining tensor sector deviates from the speed of light (i.e., $c_{13} \neq 0$), this tensor polarization will persistently contribute to, and manifest as, additional vector and scalar polarization components when evaluated in alternative, moving reference frames.

\section{Summary and Discussion}\label{sec:summary}

In this work, we investigated GW polarizations from the perspective of reference-frame transformations. 
Starting from the assumption that gravity is described by a metric theory with diffeomorphism invariance, we derived the transformation laws of all six possible GW polarizations under Lorentz boosts. 
In Sec.~\ref{sec:GWLT}, by explicitly reconstructing the metric perturbation, performing Lorentz transformations, restoring the synchronous gauge, and realigning the propagation direction, we obtained the complete polarization mixing relations between different inertial observers.

Building upon these transformation laws, in Sec.~\ref{sec:noPF}, we investigated the possible GW polarization structures in gravity theories without preferred-frame effects. 
Requiring that a propagating mode preserve its physical identity under arbitrary inertial-frame transformations imposes strong constraints on the allowed polarization combinations. 
We showed that a propagating mode carrying only tensor polarizations must propagate at the speed of light, while a purely vector mode is forbidden. 
For scalar modes, the breathing and longitudinal polarizations are not independent but must satisfy the invariant relation \(A_l/A_b = 1-k^2/\omega^2\), corresponding to a single scalar degree of freedom. 
Furthermore, a propagating mode with five physical degrees of freedom necessarily excites tensor, vector, and scalar polarizations simultaneously, with the scalar sector obeying the universal relation \(A_l/A_b=-2(1-k^2/\omega^2)\). 
The results are summarized in Sec.~\ref{sec:PolarizationsSummary}.
These results lead to a complete symmetry-based classification of GW polarizations in theories lacking preferred-frame effects, independent of the detailed form of the underlying gravitational action.
We further compared our classification with previous analyses based on EFT~\cite{Dong:2023bgt}. While the EFT approach successfully identifies several allowed polarization sectors, our reference-frame-based framework provides a more explicit characterization of the underlying propagating modes. In particular, it uniquely reveals the physical structure of the five-degree-of-freedom propagating mode and determines the corresponding scalar polarization amplitude relation \(A_l/A_b=-2(1-k^2/\omega^2)\), a feature that was not explicitly identified in previous EFT classifications.

For gravity theories with preferred-frame effects, in Sec.~\ref{sec:GWBumblebee}, we applied this framework to Bumblebee gravity, where a vector field acquires a non-vanishing vacuum expectation value and introduces a preferred reference frame. In the preferred frame, the theory propagates distinct tensor, vector, and scalar modes with different dispersion relations. We found that the scalar mode exhibits a coupled breathing-longitudinal polarization structure whose amplitudes satisfy the invariant relation \(A_l/A_b=-2(1-k^2/\omega^2)\), identical to the relation characterizing the scalar sector of a five-degree-of-freedom propagating mode in theories without preferred-frame effects. 
We further compared this result with the Einstein-Aether theory, another representative preferred-frame gravity model. In contrast to Bumblebee gravity, the scalar polarization amplitudes in Einstein-Aether theory generally do not satisfy the above relation. Therefore, the ratio \(A_l/A_b=-2(1-k^2/\omega^2)\) is not a universal feature of preferred-frame gravity theories, but instead emerges as a distinctive signature of the Bumblebee framework. 
This distinction provides a useful criterion for differentiating between alternative Lorentz-violating gravity theories through future polarization measurements.

Furthermore, once transformed to a generic observer frame, the tensor, vector, and scalar modes no longer remain pure polarization eigenstates. Instead, each mode generically excites all six GW polarizations through frame-dependent mixing.
A particularly important result concerns the vector sector. We found that vector-polarized gravitational waves in the preferred frame inevitably generate tensor polarizations for observers moving relative to that frame. The induced tensor amplitudes are proportional to the relative velocity between the observer and the preferred frame. Consequently, an observed tensor polarization does not necessarily imply the existence of a fundamental tensor propagating mode; it may instead originate from a vector mode through preferred-frame-induced polarization conversion. This provides a previously unexplored mechanism for generating tensor-like GW signals and establishes a direct observational link between vector gravitational degrees of freedom and the tensor polarizations measured by GW detectors.

Additionally, because the tensor, vector, and scalar modes generally propagate with different velocities, preferred-frame effects naturally lead to GW birefringence. In a generic observer frame, different propagating modes arrive from \emph{different directions} and at \emph{different times} while exhibiting \emph{distinct polarization mixtures}. Together, the polarization conversion and birefringence effects provide characteristic observational signatures of preferred-frame gravity theories and offer new opportunities to probe Lorentz-violating gravitational dynamics with future multi-polarization GW observations.

Overall, this work establishes a reference-frame-based framework for studying GW polarizations. The central message is that the observable polarization content of a gravitational wave cannot be completely understood without accounting for how it transforms between different inertial observers. Within this framework, we obtained a symmetry-based classification of GW polarizations in theories without preferred-frame effects, identified the physical structure of the five-degree-of-freedom propagating mode, and uncovered new observational consequences of preferred-frame gravity, including vector-to-tensor polarization conversion and GW birefringence. These results highlight the importance of observer-dependent polarization mixing in interpreting future GW observations and provide new theoretical tools for probing the fundamental structure of gravity.

\section*{Acknowledgements}

We are grateful to Hanlin Song and Zhenwei Lyu for illuminating discussions throughout the preparation of this manuscript.
This work was supported in part by the National Natural Science Foundation of China under Grant No.~12547101. HL was also supported by the start-up fund of Chongqing University under No.~0233005203009, and JZ was supported by the start-up fund of Chongqing University under No.~0233005203006.

\appendix

\section{GW Polarizations in Massive Gravity}\label{sec:GWPMG}

The action for massive gravity is formulated by the Fierz-Pauli action~\cite{Fierz:1939ix} carried by a symmetric tensor field $h_{\mu\nu}$, given by:
\begin{equation}
S=  \int d^4x \left( \mathcal{L}_{EH}^{(2)}[h] -\frac{1}{2}m^{2}(h_{\mu\nu}h^{\mu\nu}-h^{2})\right),
\end{equation}
where $\mathcal{L}_{EH}^{(2)}[h]$ represents the linearized Einstein-Hilbert Lagrangian
\begin{equation}
\begin{aligned}
    \mathcal{L}_{EH}^{(2)}[h] =& -\frac{1}{2}\partial_{\lambda}h_{\mu\nu}\partial^{\lambda}h^{\mu\nu}+\partial_{\mu}h_{\nu\lambda}\partial^{\nu}h^{\mu\lambda}\\
    &-\partial_{\mu}h^{\mu\nu}\partial_{\nu}h+\frac{1}{2}\partial_{\lambda}h\partial^{\lambda}h,
\end{aligned}
\end{equation}
and $h = \eta^{\mu\nu} h_{\mu\nu}$.
However, as we restrict our analysis to gauge-invariant actions, the standard Fierz-Pauli action is incompatible with our assumptions. We can restore this broken gauge invariance by utilizing the Stueckelberg technique~\cite{Ruegg:2003ps, Hinterbichler:2011tt, deRham:2014zqa}, wherein an extra field $A_\mu$ is introduced to accommodate the gauge degrees of freedom.
Replace $h_{\mu\nu} \to h_{\mu\nu} + \partial_\mu A_\nu + \partial_\nu A_\mu$, we obtain the following action:
\begin{equation}
S[h, A] = \int d^4x \left( \mathcal{L}_{EH}^{(2)}[h]  -\frac{1}{2}m^2 \left( H_{\mu\nu}H^{\mu\nu} - H^2 \right) \right),\label{eq:MGA}
\end{equation}
where $H_{\mu\nu} \equiv h_{\mu\nu} + \partial_\mu A_\nu + \partial_\nu A_\mu$. 
This action remains invariant under the following gauge transformations
\begin{equation}
\begin{aligned}
\delta h_{\mu\nu} &= \partial_\mu \xi_\nu + \partial_\nu \xi_\mu, \\ 
\delta A_\mu &= -\xi_\mu.
\end{aligned}
\end{equation}
By varying the gauge-invariant action \eqref{eq:MGA} with respect to the field variables, we can obtain the full set of coupled equations of motion:
\begin{equation}
\begin{gathered}
G^{(1)}_{\mu\nu}(h) - m^2(H_{\mu\nu} - \eta_{\mu\nu}H) = 0,\\
\partial_\mu F^{\mu\nu} + \partial_\mu h^{\mu\nu} - \partial^\nu h = 0,
\end{gathered}
\end{equation}
where
\begin{equation}
\begin{aligned}
G^{(1)}_{\mu\nu}(h) &=  
-\frac{1}{2} \Big[ \Box h_{\mu\nu} - \partial_\mu \partial^\alpha h_{\alpha\nu} - \partial_\nu \partial^\alpha h_{\alpha\mu} \\
&+ \partial_\mu \partial_\nu h + \eta_{\mu\nu} (\partial^\alpha \partial^\beta h_{\alpha\beta} - \Box h) \Big],\\
F_{\mu\nu} &= \partial_\mu A_\nu - \partial_\nu A_\mu. \label{eq:EoMMGA}
\end{aligned}
\end{equation}

At this stage, we can adopt the synchronous gauge to perform the GW polarization analysis, which fixes the gauge freedom by setting $h_{0i} = 0$.
We assume a GW of the form
\begin{equation}
\begin{aligned}
h_{ij} = \epsilon_{ij} \exp (i k_\mu x^\mu),\\
A_\mu = \epsilon_{i} \exp (i k_\mu x^\mu),
\end{aligned}
\end{equation}
and choose coordinates such that the wavevector is $(\omega,0,0,-k)$.
Substituting the plane-wave ansatz of the gravitational waves into the equations of motion and solving the system, we find that the wavevector must satisfy the dispersion relation $\omega^2 = k^2 + m^2$.
Furthermore, we find that the amplitudes possess only five free parameters: $\epsilon_{11}$, $\epsilon_{22}$, $\epsilon_{12}$, $\epsilon_{13}$, and $\epsilon_{23}$. The remaining components can be uniquely expressed in terms of these five parameters as follows:
\begin{equation}
\epsilon_{33} = -\frac{m^2}{\omega^2} (\epsilon_{11}+\epsilon_{22}),\label{eq:eps33}
\end{equation}
\begin{equation}
\begin{aligned}
\epsilon_{0} &= \frac{i k^2 \omega}{2m^4}  \epsilon_{33},\\
\epsilon_{1} &= \frac{i k}{m^2} \epsilon_{13}, \\
\epsilon_{2} &= \frac{i k}{m^2} \epsilon_{23}, \\
\epsilon_{3} &= \frac{i k (\omega^2+m^2)}{2m^4}  \epsilon_{33}.
\end{aligned}
\end{equation}

Evidently, \eq{eq:eps33} implies that $A_l = -\frac{2m^2}{\omega^2} A_b$, yielding an identical match with \eq{eq:lbG}.

\bibliography{refs}

\end{document}